\documentclass{aa}  
\usepackage{graphicx}
\usepackage{txfonts}
\begin{document}
   \title{Cool luminous stars: the hybrid nature of their infrared 
     spectra   
    \thanks{Tables 2 and 3 are only available in
     electronic form at the CDS via anonymous ftp to
     cdsarc.u-strasbg.fr (130.79.128.5) pub/A+A/489/1271 or via
      http://cdsweb.u-strasbg.fr/cgi-bin/qcat?J/A+A/489/1271. 
     The FTS spectra used
    in this study are also available from the same location
     (subdirectory: ftsdata/), by courtesy
    of K. H. Hinkle and S. T. Ridgway (Kitt Peak National Observatory).}
    }

   \subtitle{Carbon, oxygen, and their isotopic abundances in 23 K - M
   giant stars} 
	   
   \author{T. Tsuji}

   \offprints{T. Tsuji}

   \institute{Institute of Astronomy, School of Sciences, 
          The University of Tokyo, Mitaka, Tokyo, 181-0015 Japan\\
         \email{ttsuji@ioa.s.u-tokyo.ac.jp}
    }

   \date{Received ; accepted }

 
  \abstract
   {}
   {We determine carbon, oxygen, and their
   isotopic abundances  based on CO and OH spectra 
    in 23 red giant stars, and identify possible origin of difficulty 
    in abundance  analysis of cool luminous stars. 
   }
   {We apply the line-by-line analysis based on the
   classical micro-turbulent model and  1 D model photospheres. 
   } 
   {We found purely empirically that there is a critical value of 
    log\,${W/\nu} \approx -4.75$ ($W$ is the equivalent width and $\nu$
    the wavenumber) above which the observed lines do not follow
    the classical line formation theory based on the micro-turbulent
    model and that the classical abundance analysis can be applied 
    only to the weak lines of log\,${W/\nu} \la -4.75$. The carbon, 
    oxygen, and their isotopic 
    abundances in 23 K - M giant stars obtained from  such weak lines 
    are roughly consistent with the recent evolutionary models, although 
    the $^{12}$C/$^{13}$C puzzle remains unsolved.
    On the other hand, it is already known that the very strong lines of 
    log ${W/\nu} > -4.4$  
    are contaminated by the contribution from extra molecular layers.
    The less strong but saturated lines of $ -4.75 < {\rm log}\,{W/\nu} 
    \la -4.4 $ (intermediate-strength lines) 
    cannot be understood at all with the  abundance and turbulent velocity
    based on the weak lines. From the behavior
    shown by these lines and considering other observations such
    as the detection of H$_2$O lines, not only in the late M giants but also in
    the early M and K giants, we found that the intermediate-strength
    lines are  essentially the same as the strong lines in that they also
    include contamination from the extra molecular layers.  Thus the
    behavior of the intermediate-strength lines, including those with LEP
    as high as 2\,eV, 
    appears to be nothing but a manifestation of the warm molecular
    envelope or MOLsphere.
    }
    {The infrared spectra of K - M giant stars are a hybrid of
     at least two components originating in the photosphere and
      MOLsphere. Only the weak lines, mostly  high excitation, are 
     relatively free from the effect of MOLsphere and can be used to 
     extract photospheric abundances. The strong lines and the 
     intermediate-strength lines which actually dominate the observed 
    infrared spectra, are badly disturbed by contamination from the 
    MOLsphere. For this reason, they provide little information about
    the photosphere,  but  instead  can be new proves of the warm
    molecular envelope for which little is known yet.
     In the interpretation and analysis of the infrared spectra of
     K - M giant stars, it is essential to keep their hybrid nature
    in mind. 
     }


  \maketitle

%

\section{Introduction}

The basic principle  of stellar abundance
determination was established already about half a century
ago (e.g. Uns\"old 1955), but it is only recently that high quality 
spectroscopic data are made available for accurate abundance analyses.
Actual abundance determinations have largely been done using the 
classical methods based on  grids of model photospheres
(e.g. Gustafsson et al. 1975; Tsuji 1978; Kurucz 1979; Plez et
al. 1992). In many applications, local thermodynamical equilibrium (LTE) 
was assumed to hold (e.g. Uns\"old 1955)  and favored for its 
simplicity in  abundance determinations. The next major step was to 
treat line formation without 
the restriction of the LTE assumption (e.g. Mihalas 1978). 
More recently, the classical model photospheres describing the 
structure of the photosphere only in the radial direction (1 D) are  being
challenged by the new models considering the hydrodynamics of 
convection in 3 dimension (3 D model) (e.g. Asplund 2005).
In the regime of the coolest stars such as red giant and
supergiant stars, however, the situation may be a bit different and 
there may be another problem.

In cool giant stars representing the evolution from the red giant
branch (RGB) to the asymptotic giant branch (AGB), CNO and their 
isotopic abundances experience drastic changes  and accurate
abundance determination should be vital not only to testing the stellar 
evolutionary models but also to understanding the chemistry of 
interstellar matter which is enriched by the mass-loss from red giant stars.
Despite their importance, few abundance analyses based on high resolution
spectra have been done for cool giant stars, compared to
other type of stars. It should be remembered that high quality infrared
spectra needed for this purpose have been available, thanks to the Fourier
transform spectroscopy (FTS) pioneered by P. \& J. Connes (e.g. Connes 1970) 
and developed further in the 1970's (e.g. Maillard 1974; Hall et
al. 1979; Ridgway \& Brault 1984), even before high efficient detectors 
such as CCD innovated optical spectroscopy. Thus the 
problem here was not in the quality of the observed
data but should be due to some inherent difficulties in the spectral 
analysis of very cool stars. So far, some efforts to determine the CNO 
and their isotopic abundances were made 
for oxygen-rich case (e.g. Harris \& Lambert 1984; Smith \& Lambert
1985, 1986, 1990; Harris et al. 1985;  Tsuji 1986, 1991; 
Aoki \& Tsuji 1997) as well as for 
carbon-rich case (e.g. Lambert et al. 1984; Harris et  al.1987; 
Ohnaka \& Tsuji 1996, 1999; Abia \& Isern 1997; Ohnaka et al. 2000;
Wahlin et al. 2006; Abia et al. 2008). 
Although many interesting results were shown by these works,
we were not convinced entirely that the stellar abundance analysis  
could be done consistently for cool giant stars. 

   For example,  in our previous analysis of the CO first overtone
bands in M giant stars, we found that most lines could be interpreted
consistently with the classical model in terms of the
micro-turbulent velocity and carbon abundance (Tsuji 1986,
hereafter Paper I). The very strong
lines of low excitation could not be fitted into
this scheme but it was shown that these strong lines were contaminated
by the contributions of outer molecular layers (Tsuji 1988,
hereafter Paper II). We then analyzed the CO second overtone bands 
of the same sample of M giant stars and found that the resulting carbon
abundances appear to be larger by about
a factor of 2 compared with those based on the CO first 
overtone bands (Tsuji 1991, hereafter Paper III). 
The origin of these discrepant results remains unsolved yet. 
The difficulty in abundance analysis of very cool stars was also 
noticed by Smith \& Lambert (1990), who
found that the strong OH 1-0 and 2-0 bands gave unreasonably
large oxygen abundances compared with those based on the weaker 
bands and that strong CO lines cannot be used for abundance analysis
as well.
 
The carbon and oxygen abundances in other stars are
by no means well established yet. Even for the sun,
the classical values widely adopted (e.g. Anders \&
Grevesse 1989) had to be challenged by a new approach based
on the 3 D models, and a reduction of carbon and oxygen abundances by 
more than 50\% from the classical values was suggested (Allende Pietro 
et al. 2002). The classical values, however, were
defended by a detailed analysis of CO lines with semi-empirical model
solar photosphere by Ayres et al. (2006). These results remind us that
the accurate modeling of the photospheric structure is a 
critical issue in abundance determinations.
It is unknown if the difficulties in abundance
analysis in M giant stars outlined above are due to a limitation of
the classical 1 D model photosphere for cool giant stars or if they 
are related to other problems. For example, spectra of red giant
stars revealed such an unexpected feature as H$_2$O lines, not only
in  the early M giants and K5 giant $\alpha$ Tau (Tsuji 2001) but also in 
the earlier K giant star $\alpha$ Boo (Ryde et al. 2002). These features
cannot be fitted into our present understanding of the photospheres of
red giant stars, and suggest a new problem  in the
photosphere and/or in the atmosphere extending beyond the photosphere.
For  this reason, we analyze the
nature of the difficulty in interpreting the spectra of
cool giant stars in some detail in this paper.

\section{Input Data}

\subsection{Observed spectra}
    
     Our sample consists of one K giant and 22 M giant stars listed in 
Table 1 and shown in Fig.\,1. We use high resolution infrared spectra 
observed with the FTS
of KPNO (Hall et al. 1979). Spectra of the $H$ and $K$ band regions
are mostly from the KPNO archive, and we also observed some spectra 
of the $H$ and $K$ band regions and most spectra of the $L$ band region.
Most of these spectra were also used in our previous analyses of CO 
(Papers I, II \& III), SiO (Tsuji et al. 1994), and NH \& CN (Aoki \& 
Tsuji 1997). Some details of the spectra including the resolution, 
{\it S/N} ratio, and observing date are summarized in Table 2. We measured 
equivalent widths ({\it EW}s) of isolated lines of $^{12}$C$^{16}$O,
$^{13}$C$^{16}$O, and $^{12}$C$^{17}$O in the $H$ and  $K$ band
regions, and those of
$^{16}$OH in the $H$ and  $L$ band regions. The logarithms of
$W/\nu$ ($W$ is the measured equivalent width and $\nu$ is the
wavenumber of a line) are given in Tables 3.

\begin{table} 
\centering
\caption{  Program  stars and their  basic
stellar parameters }
\vspace{-2mm}
\begin{tabular}{ r l c c c }
\hline \hline
\noalign{\smallskip}
Obj(BS/HD) & Sp. type& $T_{\rm eff}^{~~~a}$ & $M_{\rm bol}^{~~~b}$ & 
$Mass^{~c}$ \\
          &            &  K~           & mag.             & $M_{\sun}$\\   
\noalign{\smallskip}
\hline
\noalign{\smallskip}
  $\alpha$ Tau ( 1457 ) & K5+III& 3874 & -1.7 $\pm$ 0.2 & 1.5 $\pm$ 0.3   \\
  $\delta$ Oph ( 6056 )& M0.5III & 3790 & -2.2 $\pm$ 0.3 & 1.6 $\pm$ 0.3\\
  $\nu$  Vir ( 4517 ) & M1III & 3812 & -2.2 $\pm$ 0.4 & 1.7 $\pm$ 0.4 \\
  $\alpha$ Cet (  911 ) & M1.5IIIa & 3909 & -3.2 $\pm$ 0.3 & 3.6 $\pm$ 0.4\\
  $\sigma$ Lib ( 5603 ) & M2.5III & 3596 & -3.4 $\pm$ 0.5 & 2.2 $\pm$ 0.5\\
  $\lambda$ Aqr ( 8698 )& M2.5III & 3852 & -3.4 $\pm$ 0.7 & 3.7 $\pm$ 1.2\\
  $\beta$ Peg ( 8775 ) & M2.5II-III & 3603 & -3.3 $\pm$ 0.2 & 2.2 $\pm$ 0.3\\
  $\tau^4$ Eri ( 1003 ) & M3+IIIa & 3712 & -2.9 $\pm$ 0.4 & 2.0 $\pm$ 0.4\\
  $\mu$ Gem ( 2286 ) & M3III & 3643 & -3.3 $\pm$ 0.3 & 2.3 $\pm$ 0.5\\
  $\delta$ Vir ( 4910 ) & M3III & 3643 & -2.4 $\pm$ 0.3 & 1.4 $\pm$ 0.3\\
  10 Dra ( 5226 ) & M3.5III & 3730 &  -2.9 $\pm$ 0.3 & 2.1 $\pm$ 0.4\\
  $\rho$ Per (  921 ) & M4II &    3523 & -4.1 $\pm$ 0.4 & 3.2 $\pm$ 0.5\\
  BS6861 ( 6861 )  & M4   &  3600  & -5.2 $\pm$ 2.0 &  6.3 $\pm$ 4.0\\
  $\delta^2$ Lyr ( 7139 ) & M4II & 3420 & -5.5 $\pm$ 0.8 & 5.5 $\pm$ 2.0\\
  RR UMi ( 5589 ) & M4.5III &   3397 &  -3.4 $\pm$ 0.3 & 1.6 $\pm$ 0.3\\
  $\alpha$ Her ( 6406 ) & M5Ib-II & 3293 & -5.8 $\pm$ 1.6 & 5.0 $\pm$ 2.0\\
  OP Her ( 6702 )  & M5II  &   3325 &  -4.4 $\pm$ 0.8 &  2.3 $\pm$ 1.0\\
  XY Lyr (  7009 )  & M5II  &  3300 & -5.1 $\pm$ 1.1 &  3.7 $\pm$ 1.5\\
  R Lyr ( 7157 ) &  M5III &   3313  & -4.3 $\pm$ 0.3 &  2.1 $\pm$ 0.5\\
  RZ Ari (  867 ) & M6III &  3341 &  -3.5 $\pm$ 0.6 &  1.5 $\pm$ 0.4\\
  30g Her ( 6146 ) & M6-III &  3298 & -4.2 $\pm$ 0.4 & 2.0 $\pm$ 0.6\\
  SW Vir (114961) & M7-III: & 2886 &  -4.4 $\pm$ 0.9 &  1.3 $\pm$ 0.3\\
  RX Boo (126327) & M7.5-8 &  2850 & -4.6 $\pm$ 0.8 &  1.3 $\pm$ 0.3\\
\noalign{\smallskip}
\hline \hline
\noalign{\smallskip}
\end{tabular}
\vspace{-3mm}
\begin{list}{}{}
\item[$^{\mathrm{a}}$]  based on the infrared flux method. The
			uncertainty of $T_{\rm eff}$ for individual
                        object is estimated to be $\pm100$\,K. 
\item[$^{\mathrm{b}}$]  based on the Hipparcos trigonometric parallaxes
                        and measured bolometric fluxes.
\item[$^{\mathrm{c}}$]  based on the evolutionary tracks of Fig.\,1.  
\end{list}
\vspace{-2mm}
\end{table}

\onltab{2}{
\begin{table*}
\caption{Observed spectra}
\begin{tabular}{lccccc}
obj.   & Filter & res.(mK) & S/N & Date of obs. &  FTS sp. in ftsdata/\\
-------&--------&----------&-----&--------------&--------------------\\
alf Tau &   H   &     46.19 &    60 &  Aug. 17, 1976 &  kp0222\\   
alf Tau &  K    &    46.19   &  76   & Sep. 28, 1976  &  kp0226\\      
alf Tau  &  L    &     49.90  &  159  &  Oct. 21, 1977 & kp0628\\
del Oph &  H,K  &     49.70 &   75  & Jun. 24, 1977  & kp0201\\
nu Vir  &  H    &     41.21 &   51  & Apr. 08, 1982  & kp0321\\
nu Vir  &  K    &    41.20  & 115   & Apr. 09, 1982  & kp0332\\
alf Cet &  H,K  &    46.19  &  69   & Sep. 28, 1976  & kp0221\\
sig Lib &  H,K  &    46.19  &  89   & Aug. 18, 1976  & kp0228\\
lam Aqr &  H,K  &    46.19  &  61   & Jul. 06, 1976  & kp0313\\
bet Peg &  H,K  &    48.76  & 103  & May  12, 1976  & kp0104\\
tau4 Eri&  H,K  &    46.19  &  59  & Sep. 29, 1976  & kp0127\\
mu Gem	&  H    &    46.19  &   -  & Nov. 07, 1976  & kp0219\\
mu Gem  &  K    &    16.53  &  80  & Jan. 27, 1988  & kp0902\\
del Vir &  H    &    46.19  &  54  & Jun. 18, 1976  & kp0212\\
del Vir &  H,K  &    49.70  &  86  & Jun. 24, 1977  & kp0207\\
10 Dra  &  H,K   &   49.82 &   61 &  Jun. 28, 1977 &  kp0131\\
rho Per &  H   &     29.70 &   44 &  Apr. 07, 1982 &  kp0317\\
rho Per &  K     &   22.28 &   66 &  Apr. 08, 1982 &  kp0329\\
BS6861  &  H,K  &    46.19 &   42 &  Sep. 29, 1976 &  kp0114\\
del2 Lyr&  H,K   &   46.19  &  78 &  Apr. 11, 1976 &  kp0111\\
RR UMi  &   H,K  &    46.19 &   61 &   Aug. 18, 1976 & kp0227\\
alf Her &  H,K  &    49.70 &  106 &  Jun. 24, 1977 &  kp0202\\
alf Her &  L    &    16.53 &   45 &  Jul. 09, 1987 &  kp0940\\
OP Her  &  K    &    16.53 &  142 &  Jul. 01, 1988 &  kp1124\\
XY Lyr  &  K    &    16.53 &   86 &  Jul. 08, 1987 &  kp0930\\
R Lyr   &  H     &   29.70  &  70 &  Apr. 08, 1982 &  kp0325\\
R Lyr   &  K    &    16.53  &  65 &  Jul. 07, 1987 &  kp0923\\
R Lyr   &  L     &   16.53 &  107 &  Jul. 09, 1987 &  kp0941\\
RZ Ari  &  H,K   &   46.19 &   88 &  Aug. 17, 1976 &  kp0115\\
RZ Ari  &  L      &  16.53 &   65 &  Oct. 02, 1987 &  kp1028\\
30g Her &  H,K  &    49.10 &  102 &  Jun. 30, 1977 &  kp0128\\
30g Her &  L     &   16.53  &   - &  Jul. 09, 1987 &  kp0939\\
SW Vir  &  H,K   &   46.19 &   65 &  Jun. 18, 1976 &  kp0120\\
SW Vir  &  L      &  16.53  &  88 &  Jul. 09, 1987 &  kp0936\\
RX Boo	&  H,K   &   48.10  &  62 &  Jan. 14, 1976 &  kp0105\\
RX Boo  &  L      &  16.53 &  108 &  Jul. 09, 1987 &  kp0937\\
alf CMa &  H      &  29.70 &   -  &  Apr. 07, 1982 &  kp0318\\
alf CMa &  K      &  22.28 &   -  &  Apr. 09, 1982 &  kp0330\\
IRC+10216 &  L      &  16.53 &   -  &  Oct. 02, 1987 &  kp1033\\ 
\end{tabular}
\end{table*}
}

\onltab{3}{
\begin{table*}
\caption{Measured equivalent width data (only a few lines at the
 beginning of  table3  as examples)}
\begin{tabular}{llcccccc}
 star      & isotope & v'  v'' & rot. tr. & nu(cm-1)  & log gf & L.E.P.(cm-1) & logEW/nu \\
  ----------&--------&--------&----------&-----------&--------&--------------&---------\\
alf Tau & C12O16 &  3   1 &  P   2 &  4199.480 & -6.282 & 2154.701 & -4.739 \\
alf Tau & C12O16 &  2   0 &  P   9 &  4222.953 & -6.128 & 172.978 & -4.585 \\
alf Tau & C12O16 &  4   2 &  R  68 &  4237.805 & -4.241 & 12982.145 & -4.721 \\
alf Tau & C12O16 &  3   1 &  R   8 &  4238.289 & -5.600 &  2280.402 & -4.611\\ 
alf Tau & C12O16 &  4   2 &  R  36 &  4243.637 & -4.604  &  6763.398 & -4.608\\
alf Tau & C12O16 &  3   1 &  R  18 &  4266.090 & -5.248  &  2794.075 & -4.572\\
alf Tau & C12O16 &  3   1 &  R  22 &  4275.180 & -5.154  &  3105.648 & -4.553\\
alf Tau & C12O16 &  2   0 &  R  92 &  4292.047 & -4.817 &  16004.551 & -5.067\\
alf Tau & C12O16 &  3   1 &  R  32 &  4292.750 & -4.969 &   4148.160 & -4.563\\
alf Tau & C12O16 &  3   1 &  R  35 &  4296.566 & -4.923 &   4533.898 & -4.595\\
\noalign{\smallskip}
\end{tabular}
\end{table*}
}

\begin{figure}
\centering
\includegraphics[width=8.5cm]{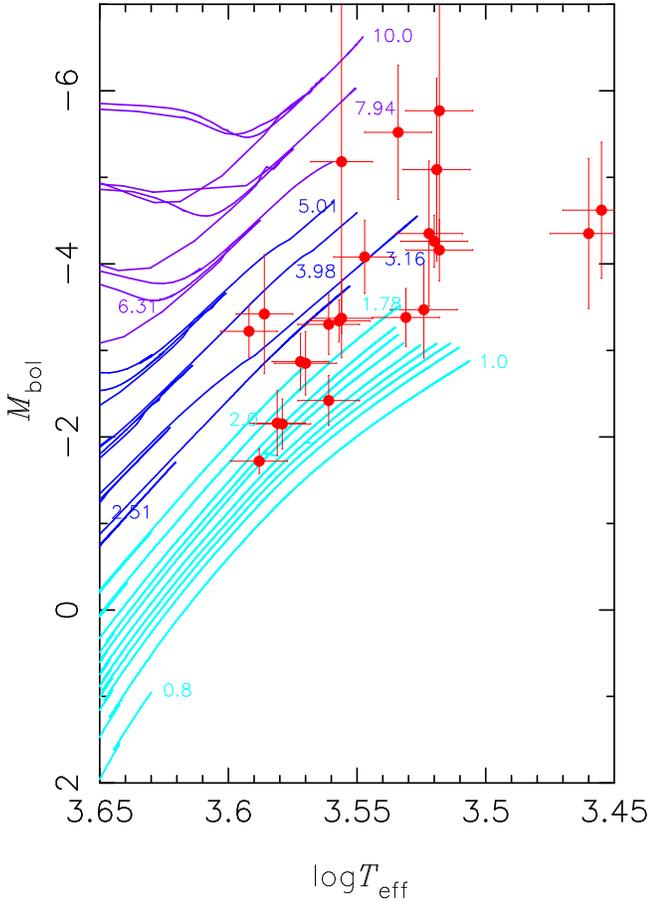}
\caption{
Program stars shown by  filled circles on the HR diagram 
and  recent evolutionary tracks (Claret 2004) shown by solid lines.
Stellar masses (in unit of solar mass $M_{\sun}$) are indicated 
on some evolutionary tracks.
}
\label{Fig1.eps}
\end{figure}

\begin{table} 
\hspace{-20mm}
\caption{CNO abundances used in model photospheres}
\vspace{-2mm}
\begin{tabular}{ l c c l}
\hline \hline
\noalign{\smallskip}
\noalign{\smallskip}
   & case a & case b  \\    
\noalign{\smallskip}
\hline
\noalign{\smallskip}
 log $A_{\rm C}^{~~a} $    &  8.35      &   8.08      \\
 log $A_{\rm N}^{~~b}$    &   8.11     &    8.45     \\
 log $A_{\rm O}^{~~c}$    &   8.69      &   8.69      \\
\noalign{\smallskip}
\hline \hline
\noalign{\smallskip}
\end{tabular}
\vspace{-3mm}
\begin{list}{}{}
\item[$^{\mathrm{a}}$]   Tsuji (1991). 
\item[$^{\mathrm{b}}$]   Aoki \& Tsuji (1997).
\item[$^{\mathrm{c}}$]   Allende Prieto et al. (2002).  
\end{list}
\vspace{-3mm}
\end{table}

\begin{table} 
\centering
\caption{  Model photospheres}
\vspace{-2mm}
\begin{tabular}{ c c c r r c c c}
\hline \hline
\noalign{\smallskip}
\noalign{\smallskip}
 case   & $ R$ & $ T_{\rm eff} $ & log\,$g$  &
$T_{0}^{~a}$ & $d^{~b}$ & log $F_{\rm L}^{~c}$  & log $R_{\rm L}^{~d}$ \\  
\noalign{\smallskip}
     & ~$R_{\sun}$    &  K~~ &   & K~~   &    &    &  \\
\noalign{\smallskip}
\hline
\noalign{\smallskip}
\noalign{\smallskip}
$M =$ & $M_{\sun}$   &   &        &         &        &         &  \\
\noalign{\smallskip}
  a     &    ~50 &   4000 & 1.04 &   2196 &   1.06 &  12.206 &   -2.542 \\ 
   a    &    ~50 &   3900 & 1.04 &   2185 &   1.08 &  12.193 &   -2.572 \\
   a    &    ~75 &   3800 & 0.69 &   2117 &   1.13 &  12.185 &   -2.609 \\  
   a    &    ~75 &   3700 & 0.69 &   2094 &   1.13 &  12.171 &   -2.641 \\
   a    &    100 &   3600 & 0.44 &   2023  &  1.18 &  12.160  & -2.679  \\
\noalign{\smallskip}
   b     &    100 &   3600 & 0.44 &   2074 &   1.18 &  12.159 &   -2.677 \\
   b     &    100 &   3500 & 0.44 &   2037 &   1.18 &  12.144 &   -2.711 \\  
   b    &    150 &   3400 &  0.09 &  1857 &   1.30 &  12.137 &   -2.755 \\
   b     &    150 &   3300 & 0.09 &   1628 &   1.29 &  12.118 &   -2.787 \\
   b    &    200 &   3200 & -0.16 &   1396 &   1.40 &  12.102 &   -2.825 \\
\noalign{\smallskip}
   b    &    200 &   3100 & -0.16 &   1291 &   1.38 &  12.064 &   -2.842 \\ 
   b     &    250 &   3000 & -0.36 &   1079 &   1.51 &  12.031 &   -2.866 \\
   b    &    250 &   2900 & -0.36 &   1021 &   1.47 &  11.974 &   -2.868 \\ 
   b     &    300 &   2800 & -0.52 &    895 &   1.58 &  11.931 &   -2.886 \\ 
\noalign{\smallskip}
\noalign{\smallskip}
$M =$ & $2\,M_{\sun}$   &   &        &         &        &         &  \\
\noalign{\smallskip}
   a     &    ~50 &   4000 & 1.34 &  2251 &   1.04 &  12.200 &   -2.536 \\ 
   a    &    ~50 &   3900 &  1.34 & 2234 &   1.04 &  12.187 &   -2.566 \\
   a     &    ~75 &   3800 & 0.99 &  2172 &   1.06 &  12.178 &   -2.602 \\
   a     &    ~75 &   3700 & 0.99 &  2152 &   1.06 &  12.163 &   -2.634 \\
   a    &    100 &   3600 & 0.74 &  2089 &   1.08  &  12.152 &   -2.670 \\
\noalign{\smallskip}
   b    &    100 &   3600 & 0.74 &  2142 &   1.08 &  12.151 &   -2.668 \\
   b     &    100 &   3500 & 0.74 &  2107 &   1.08 &  12.135 &   -2.702 \\   
   b    &    150 &   3400 & 0.39 &  2003 &   1.13 &  12.124 &   -2.742 \\
   b    &    150 &   3300 & 0.39 &  1884 &   1.12 &  12.104 &   -2.773 \\
   b     &    200 &   3200 & 0.14 &  1652 &   1.17 &  12.083 &   -2.806 \\
\noalign{\smallskip}
   b     &    200 &   3100 & 0.14 &  1525 &   1.16 &  12.041 &   -2.819 \\
   b    &    250 &   3000 & -0.06 &   1342 &   1.20 &  11.998 &   -2.833 \\ 
   b    &    250 &   2900 & -0.06 &   1234 &   1.19 &  11.936 &   -2.830 \\
   b     &    300 &   2800 & -0.22 &  1102 &   1.22 &  11.882 &   -2.837 \\
\noalign{\smallskip}
\hline \hline
\noalign{\smallskip}
\end{tabular}
\vspace{-3mm}
\begin{list}{}{}
\item[$^{\mathrm{a}}$]   temperature at $\tau_0 = 10^{-6}$ where $\tau_0$
			 is the optical depth defined by the continuous
			 opacity at $\lambda = 0.81 \mu$m.
\item[$^{\mathrm{b}}$]   $ d = r(\tau_0 = 10^{-6})/R $ is a measure
			 of the extension of the photosphere.
\item[$^{\mathrm{c}}$]  $F_{L}$ is the flux in the $L$ band in unit of
                       erg\,cm$^{-2}$\,s$^{-1}$ per $\Delta\,\lambda = 1$\,cm.
\item[$^{\mathrm{d}}$]  $R_{L} = F_{\rm bol}/F_L$  where $ F_{\rm bol}$
			 is the bolometric flux.                        
\end{list}
\vspace{-3mm}
\end{table}

\subsection{Model photospheres}
   We apply classical 1 D model photospheres that are essentially the 
same as those adopted in our previous works (Tsuji 1978; Paper I - III), 
except that we now assume  that the photosphere is spherically 
symmetric (SS) rather than plane-parallel (PP) and some molecular 
opacity data are updated (Tsuji 2002). We include 34 elements in the 
solar composition (Table 1 of Tsuji 2002) except for C, N, and O 
for which we assume two cases a and b listed in Table 4
\footnote{Abundances are expressed throughout on the scale of 
log\,$A_{\element[][]{H}} = 12.0 $. }.  Case a represents the
 mean C  and N abundances of early M giants and case b of
late M giants based on our previous analyses (Paper III; Aoki \& Tsuji
1997). The masses, radii, effective temperatures, and log\,$g$ of our model 
photospheres are listed in Table 5
\footnote{We refer to a model by its
case/mass (in $M_{\sun}$)/radius (in $R_{\sun}$) /$T_\mathrm{eff}$ (in K). 
For example, the first model
photosphere in Table 5 is referred to as a/1.0/50/4000.
}. 
We assume masses to be one and two solar masses to see the
effect of mass. However, we used the models with $ M = 2 M_{\sun}$ in the
following analyses in this paper, since this appears to be closer
to the masses of most red giant stars we are to study (Table 1). 
Also, the changes in the photospheric structure are rather minor if
the masses are increased beyond 2\,$M_{\sun}$, since the differences
between the SS and PP models tend to be reduced for larger masses. 
We assumed the micro-turbulent velocity in modeling to be 3 km\,s$^{-1}$ 
throughout.

   Some characteristics of the resulting models are given in Table 5:
The surface temperature $T_0$ at $\tau_0 = 10^{-6}$ ($\tau_0$ is the
optical depth defined by the continuous opacity at $\lambda = 0.81
\mu$m) and  extension parameter defined by $ d = r(\tau_0 = 10^{-6})/R$.
Some of our models are compared with those by Plez (1992)
in Fig.2. These models are based on the similar assumptions except that
Plez et al.(1992) applied the opacity sampling method while we use 
the band model method in taking the blanketing effect of molecular lines
into account. The agreements are generally fair apart from at 
the very surface of the coolest model. 
In evaluating the flux, we calculate spectra at a sampling interval 
of 0.05\,cm$^{-1}$ using a detailed line list including CO (Guelachivili 
et al. 1983; Chackerian \& Tipping 1983), OH (Jacquinet-Husson et
al.1999), CN (Cerny et al. 1978; Bausclicher et al. 1988), H$_2$O 
(Partridge \& Schwenke 1997), and SiO (Lavas et al. 1981; Tipping \& 
Chackerian 1981).   Based on the resulting spectra, the flux
in the $L$ band of the wide-band photometry, $F_L$, and log $R_{L}$, 
where $R_{L} = F_{\rm bol}/F_L$ with the bolometric flux
$F_{\rm bol} = \sigma\,T_{\rm eff}^4/\pi$ ($\sigma$: Stefan-Boltzmann
constant), are obtained (Table 5).

\begin{figure}
\centering
\includegraphics[width=8.0cm]{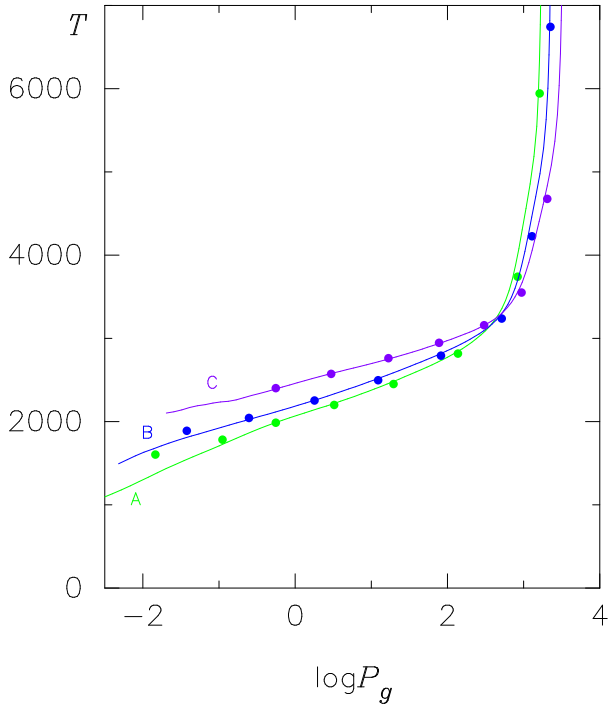}
\caption{
Some  of our model photospheres (solid lines) are compared with those by 
Plez et al.(1992) (dots). $ M/M_{\sun}$, $T_{\rm eff}$, log\,$g$, and 
$R/R_{\sun}$ are as follows. A: 1.0/2750/-0.50/295,  B:
 1.0/3200/0.00/166, and C: 2.0/3600/0.50/132. 
}
\label{Fig2.eps}
\end{figure}

The stellar mass has a significant effects on the surface 
temperature $T_0$ and the extension parameter $ d $ especially 
at low values of 
$ T_{\rm eff} $, and we will examine its effects on abundance 
determination (Sects.\,4.3 \& 5.3). The calculated 
values of log\,$R_L$  differ only slightly, however, from our
previous values (Tsuji 1981), which would be expected because the $L$ band 
region is relatively free from the heavy line-blanketing effect.
For $T_\mathrm{eff} = 3600$\,K, we compare the results for cases a
and b, but the differences are rather minor (Table 5).

\subsection{Basic stellar parameters}

For our sample of  23 red giant stars, 
the effective temperatures  are determined by applying the
infrared flux method (Blackwell et al. 1980) using the log $R_{L}$ 
values from Table 5 and the observed values of log $R_{L}$ (Tsuji
1981).   The resulting values of $T_{\rm eff}$ agree with our previous
results (Tsuji 1981) in general to within 50\,K. Our
$T_{\rm eff}$ scale is also consistent with the empirical ones based on the
measured angular diameters by Ridgway et al. (1980) and  more
recent results (e.g. Dyck et al. 1998; Perrin et al. 1998; van Belle et
al.1999; Mondal \& Chandrasekhar 2005).  We also found that our effective
temperatures of individual objects agree with the empirical values of these 
authors, in most cases, to within about 100\,K; we therefore assume 
an error bar of 100\,K for our estimations of $T_{\rm eff}$. 
With the Hipparcos parallaxes 
(ESA 1997)\footnote{This research has made use of the VizieR catalogue
access tool, CDS, Strasburg, France, via ADC/NAOJ.} 
and the
bolometric luminosities obtained by the integration of the SEDs (Tsuji
1981), the absolute bolometric magnitudes are determined.
The uncertainty in $M_{\rm bol}$ is dominated by errors in the parallaxes. 
The errors in the photometric data are estimated to be about $\pm 0.1$
mag, including absolute calibration errors of a few percent (Bersaneli 
et al. 1991).

Our sample is shown on the HR diagram (Fig.\,1) and an evolutionary mass
is estimated by comparison  with the recent evolutionary tracks by 
Claret (2004). Unfortunately, nearly half of our sample cannot be 
accounted for by the models of Claret, and we had to extrapolate the 
evolutionary tracks to estimate stellar masses given in Table 1. The 
error bars of masses are derived from those of $M_{\rm bol}$ and 
$T_{\rm eff}$.  For the objects with $T_{\rm eff}$ lower than
$\approx 3000$\,K, the extrapolations needed are so significant that the
uncertainties should be much larger, and we hope that the theoretical 
evolutionary tracks can be extended to account for these objects.

\section{Method of Analysis}

Given a model photosphere characterized by a set of mass, radius,
and luminosity (or reduced to  $T_{\rm eff}$ and log\,$g$ 
in PP models), a given {\it EW}
can be interpreted in terms of abundance and micro-turbulent
velocity, within the framework of the classical theory of line 
formation. Actually abundances of 34 elements are assumed in
our model photosphere, and what we are to determine is corrections
to the abundances assumed. 
In our analysis, we apply the line-by-line analysis based on the
classical micro-turbulent model.
In this method, we plot the abundance correction $\Delta$log\,$A$
obtained for an assumed value of $\xi_{\rm micro}$ against the 
log\,$W/\nu$ of the line. We expect that 
larger (smaller) abundance corrections are needed if smaller (larger) values 
of the micro-turbulent velocity are assumed especially for more
saturated lines, while abundance corrections are independent of the assumed 
$\xi_{\rm micro}$ value for very weak lines free from  saturation effect.    
The micro-turbulent velocity $\xi_{\rm micro}$ is determined so that 
abundance corrections for individual lines show the least dependence  on 
{\it EW}s (as for detail, see Paper I, Sect.\,5.1). 
 
Also, the probable errors (PEs)
of the resulting abundance (correction) and micro-turbulent velocity  
are essentially determined from the scatter in the data points plotted on
the $\Delta$log\,$A$ - log\,$W/\nu$ plane, as explained in detail
in Paper I (see its Fig.\,2 and related explanation). Accordingly, the probable
errors are robust assessments of the internal errors of our analysis and 
any result that has a large PE should be assumed to have a large
uncertainty (e.g., some $\xi_{\rm micro}$ values with large PEs in Table 6). 
Apart from the internal error, we notice that there should be
a systematic error due to a selection of the lines.
The analysis, which is based on the classical micro-turbulent model 
outlined above, should be applied only to lines weaker than a 
critical value, since otherwise a large systematic error will 
be introduced,  as will be shown in Sects.\,4 \& 5.

\section{Reanalysis of CO}

We analyzed the lines of the CO first and second overtone bands
separately in Papers I and III, respectively, and we 
summarize briefly the difficulties in these analyses. We then
reanalyze the CO lines with the same observed and spectroscopic data 
as before, but treat the first and second overtone bands together. 

\subsection{Previous analysis of CO}

The method outlined in Sect.3 has been applied to the lines of the 
$^{12}$C$^{16}$O first overtone bands (see Figs.\,1a-g of Paper I).
The analysis in Paper I was done using PP model 
photospheres, but we confirmed that the results are essentially the same 
for SS model photospheres of the same $T_{\rm eff}$. As an example,
we repeat the line-by-line analysis of $\alpha$ Her in Fig.\,3a: For 
assumed values of $\xi_{\rm micro}$ = 3.0, 3.5, and 4.0 km\,s$^{-1}$,
 the CO lines of $ {\rm log}\,W/\nu \la -4.5$ (those shown by filled
circles) show  behavior expected from the classical micro-turbulent 
model outlined in Sect.\,3, and it is possible to find a
$\xi_{\rm micro}$ value for which the abundance corrections do not
depend on $ {\rm log}\,W/\nu$ values. The resulting values of 
$\xi_{\rm micro} = 3.81 \pm 0.43$ km\,s$^{-1}$ 
and log\,$A_{\rm C} = 8.08 \pm 0.27$
in fact provide a null mean logarithmic abundance correction as confirmed  
in Fig.\,3b
\footnote{ Some differences with the result for $\alpha$ Her in Table 4
of Paper I ($\xi_{\rm micro} = 3.30 \pm 0.38$ km\,s$^{-1}$ and 
log\,$A_{\rm C} = 7.93 \pm 0.28$), may be due to the use of SS 
instead of PP models  in the present analysis. Thus the effect of
sphericity is  certainly not negligible but not very important.}.

Although it appeared to be possible
that a self-consistent abundance analysis could be done  
for the lines of log $W/\nu \la -4.5$,
 it was found that the lines of log $W/\nu > -4.5$
 could not be fitted into the above scheme, 
but behave quite differently (also see Figs.\,1 and 2 of Paper II). 
 From a detailed analysis of the line profiles, strengths, and velocity 
shifts of these strong lines, we suggested  that these lines should be
contaminated by the contributions of extra molecular layers
above the photosphere as detailed in Paper II and hence  
these lines cannot be used for abundance analysis. 

We then analyzed the lines of the CO second overtone bands by the
same method as for the CO first overtone bands (Paper III).  
In this case, many weak lines were available and the abundances
determined were relatively insensitive to the value of $\xi_{\rm
micro}$. At the same time, the value of $\xi_{\rm micro}$ could not be 
specified well for some cases (Fig.\,2 of Paper III). The resulting carbon
abundances based on the CO second overtone bands, however, appeared
to be about twice as high of those based on the CO first overtone bands.
The origin of such a systematic difference between the
carbon abundances based on the CO first and second overtone bands
is difficult to understand, since both the analyses of the first and
second overtone bands have been done internally self-consistent. 

In view of a general preference of weak lines in abundance analysis,
it may be possible to adopt the carbon abundances based on the
CO second overtone bands and abandon the results based on the
first overtone bands. However, the CO first overtone
bands are useful for investigating isotopic abundances, since
$^{13}$C$^{16}$O and $^{12}$C$^{17}$O lines of the first overtone
bands are good indicators of $^{13}$C and $^{17}$O abundances,
respectively (Sect.\,6). We should then investigate in some detail
the problem with the first overtone bands  if we are to use them 
for investigating the isotopic abundances. We    
assumed that the isotopic ratios can be determined even if 
the elemental abundances themselves cannot be determined accurately
in our preliminary analysis of the isotopic ratios (Tsuji
2007), but we think it necessary to  reconsider  such a viewpoint.

\begin{figure}
\centering
\includegraphics[width=8.0cm]{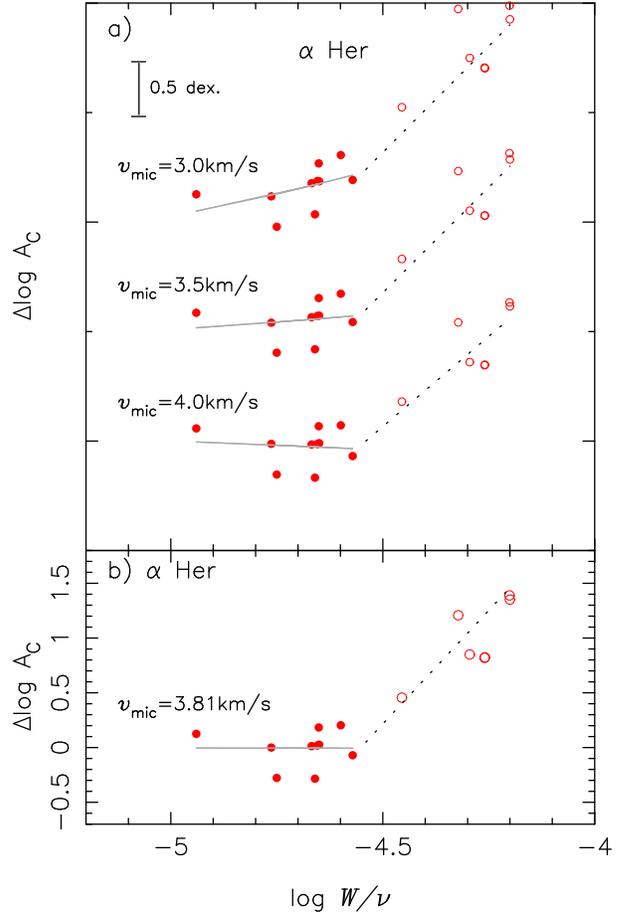}
\caption{
 {\bf a)} Logarithmic abundance corrections  for the lines of 
the CO first overtone bands of $\alpha$ Her plotted against  
the observed  log\,$W/\nu$ values for assumed values of 
$\xi_{\rm micro}$ = 3.0, 3.5, and 4.0 km\,s$^{-1}$.  
The analysis outlined in Sect.3 is applied to the lines of 
log\,$W/\nu < -4.5 $ (filled circles) with our model b/2.0/150/3200. {\bf b)}  
Confirmation of the null logarithmic  abundance corrections  for 
log\,$A_{\rm C}$ = 8.08  and $\xi_{\rm micro}$ = 3.81 km\,s$^{-1}$, which
are the solution of the line-by-line analysis of $\alpha$ Her in a). 
}
\label{Fig3.eps}
\end{figure}

\begin{figure}[ht]
\centering
\includegraphics[width=8.0cm]{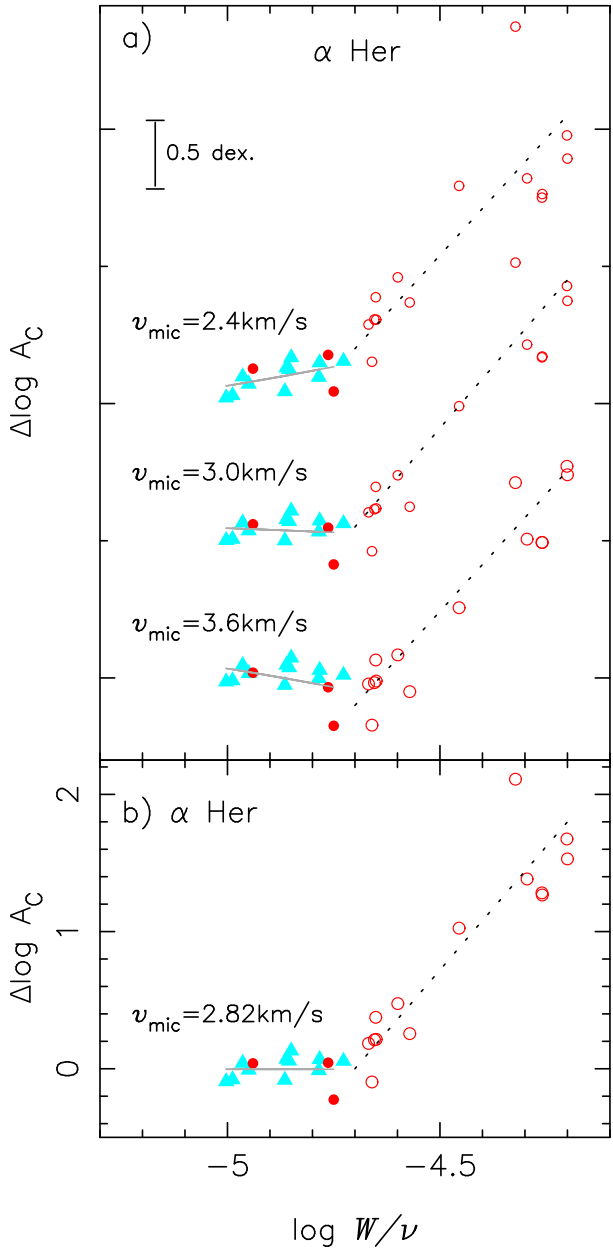}
\caption{
{\bf a)} Logarithmic abundance corrections  for the lines of the CO
first and second overtone bands  in  $\alpha$ Her plotted against
the observed $ {\rm log} W/\nu $ values for assumed values of
$\xi_{\rm micro}$ = 2.4, 3.0, and 3.6 km\,s$^{-1}$. The CO lines
of the first and second overtone bands are shown by  circles
and triangles, respectively (model photosphere: b/2.0/150/3300).
{\bf b)}  Confirmation of the null logarithmic  abundance corrections  for
log\,$A_{\rm C}$ = 8.40  and $\xi_{\rm micro}$ = 2.82 km\,s$^{-1}$, which
are the solution of the line-by-line analysis of the weak lines (filled
symbols) in a).
}
\label{Fig4.eps}
\end{figure}

\begin{figure}
\centering
\includegraphics[width=8.0cm]{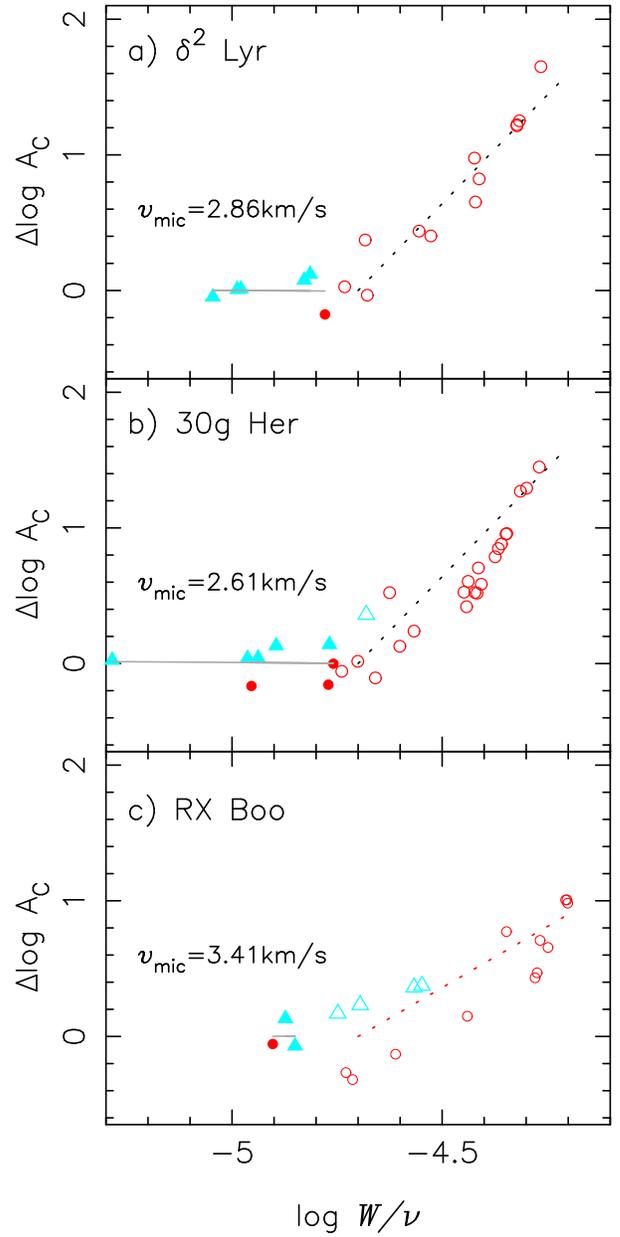}
\caption{
{\bf a)} Confirmation of the null logarithmic  abundance corrections
for log\,$A_{\rm C}$  = 8.22 and $\xi_{\rm micro}$ = 2.86\,km\,s$^{-1}$,
resulting from the line-by-line analysis of the weak lines of CO in
$\delta^{2}$ Lyr (model: b/2.0/150/3400).
{\bf b)} The same as a) but for  log\,$A_{\rm C}$ = 8.35 and
$\xi_{\rm micro}$ = 2.61\,km\,s$^{-1}$ in 30g Her (model: b/2.0/150/3300).
{\bf c)} The same as a) but  for  log\,$A_{\rm C}$ = 8.32
and $\xi_{\rm micro}$ = 3.41\,km\,s$^{-1}$ in RX Boo (model: b/2.0/250/2900).
}
\label{Fig5.eps}
\end{figure}

\subsection{The first and second overtone bands}

Although OH analysis to be detailed in Sect.\,5 suggests that the 
lines stronger than a
critical value of $ {\rm log} W/\nu \approx -4.75$ behave differently
against the weaker lines, we are unable  to see this effect in the CO 
lines of Fig.\,3, since only a few lines of the CO first overtone
bands with  $ {\rm log}\,W/\nu \la -4.75$ could be measured especially in 
the later M giant stars (e.g. only 3 lines could be measured in $\alpha$
Her shown in Fig.\,3). To include more weak lines,  
we tried to analyze together the first and second overtone bands of CO.
The results for $\xi_{\rm micro}$ = 2.4, 3.0 and 3.6 km/s 
are shown in Fig.4a, where the lines of the first and second overtone 
bands of CO are shown by circles and triangles, respectively. 
All these lines are analyzed together, regardless whether 
 they are of the first or of the second overtone bands.    
Unlike Fig.\,3a based on the CO lines of the first overtone bands alone,
it now appears that there is a clear distinction between the lines of 
$ {\rm log}\,W/\nu > -4.75$ (shown by the open symbols) and those of 
$ {\rm log}\,W/\nu \la -4.75$ (shown by the filled symbols).
Only the lines of $ {\rm log}\,W/\nu \la -4.75$ show the behavior
expected from the classical micro-turbulent model of line formation
theory noted in Sect.\,3.
 
\begin{figure}
\centering
\includegraphics[width=8.0cm]{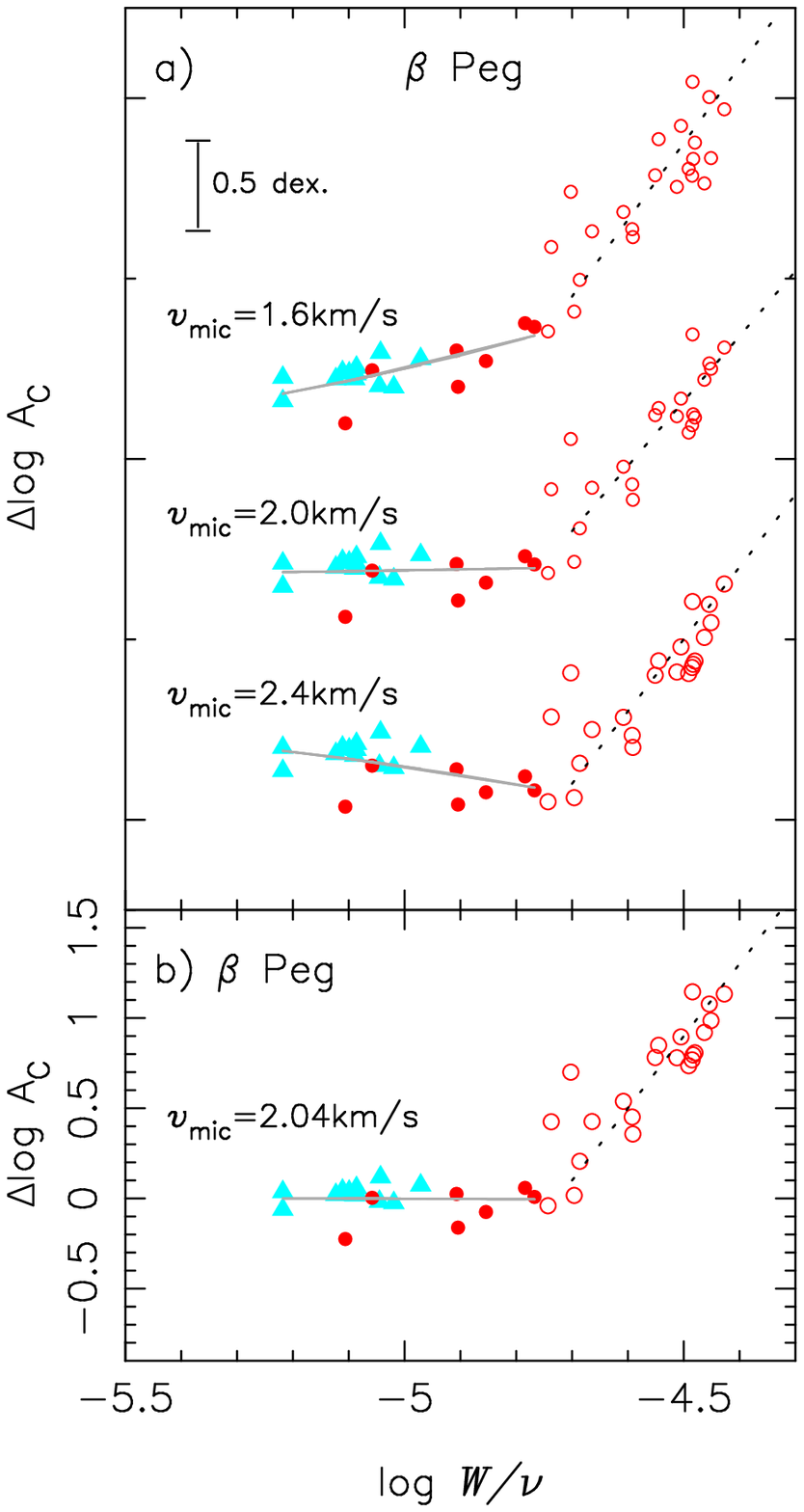}
\caption{
{\bf a)} Logarithmic abundance corrections  for the lines of the CO
first and second overtone bands in $\beta$ Peg plotted against
the observed $ {\rm log} W/\nu $ values for assumed values of
$\xi_{\rm micro}$ = 1.6, 2.0, and 2.4\,km\,s$^{-1}$ (model: a/2.0/100/3600).
{\bf b)}  Confirmation of the null logarithmic  abundance corrections  for
log\,$A_{\rm C}$ = 8.27  and $\xi_{\rm micro}$ = 2.04\,km\,s$^{-1}$, which
are the solution of the line-by-line analysis of the weak lines (filled
symbols) in a).
}
\label{Fig6.eps}
\end{figure}

\begin{figure}
\centering
\includegraphics[width=8.0cm]{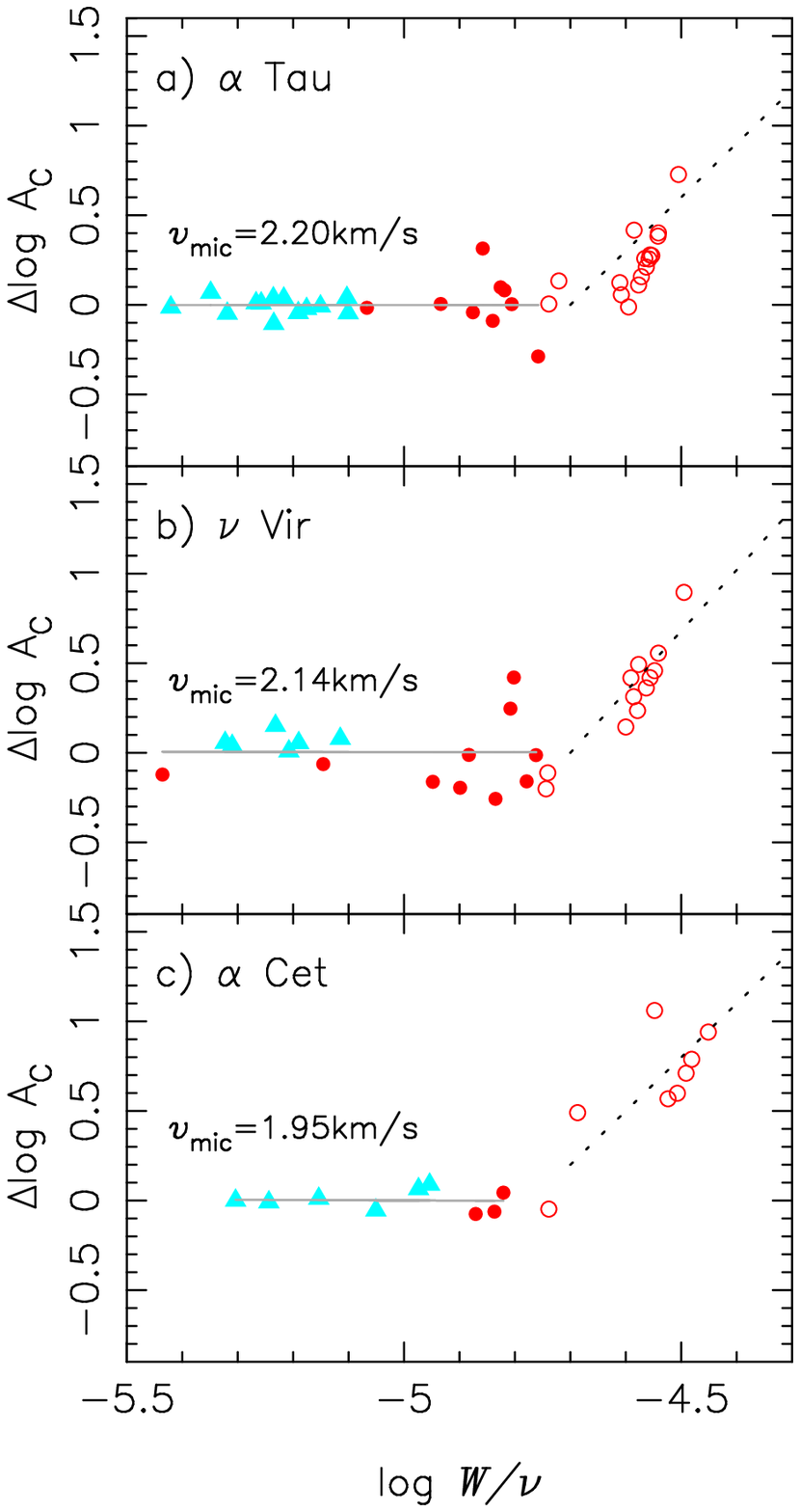}
\caption{
{\bf a)} Confirmation of the null logarithmic  abundance corrections
for log\,$A_{\rm C}$ = 8.38 and $\xi_{\rm micro}$ = 2.20\,km\,s$^{-1}$,
resulting from the line-by-line analysis of the weak lines of CO in
$\alpha$ Tau (model: a/2.0/50/3900).
{\bf b)} The same as a) but for  log\,$A_{\rm C}$ = 8.13 and
$\xi_{\rm micro}$ = 2.14\,km\,s$^{-1}$ in $\nu$ Vir (model: a/2.0/75/3800).
{\bf c)} The same as a) but for log\,$A_{\rm C}$ = 8.64 and
$\xi_{\rm micro}$ = 1.95\,km\,s$^{-1}$ in $\alpha$ Cet (model: a/2.0/50/3900).
}
\label{Fig7.eps}
\end{figure}

It is evident in Fig.\,4a  that it is impossible to find a turbulent 
velocity that provides a consistent abundance correction to all 
the lines measured covering $ -5.0 < {\rm log}\,W/\nu < -4.2$ . Thus, 
the critical value of $ {\rm log} W/\nu \approx -4.75$ found for the CO 
lines is purely empirical as for OH to be discussed in Sect.\,5.1. 
 Unfortunately, we thought previously (Paper I) that
such a critical value is $ {\rm log} W/\nu \approx -4.5$ rather than
-4.75. In this paper, we confirm that this new critical value of
$ {\rm log} W/\nu \approx -4.75$ 
should apply to all red giant stars that we are to study\footnote{
With the critical value that discriminates the lines well
interpreted by the classical line formation theory from those that
cannot be, we distinguish three groups of lines to be referred to as; 
1) the weak lines (log\,$W/\nu  \la -4.75$, although  
these lines are not necessarily very weak lines free from saturation
effect). 2) the intermediate-strength lines ($ -4.75 < {\rm log}\,W/\nu 
\la -4.4$), and 3)  the strong lines (log\,$W/\nu  > -4.4$, these lines
are  mostly low excitation lines of CO discussed in Paper II). 
}.

From the weak lines which behave as expected for
different values of $\xi_{\rm micro}$ in Fig.\,4a, we can determine 
$\xi_{\rm micro}$ value using the method described in Sect.\,3. The solution 
is found to be $\xi_{\rm micro} = 2.82 \pm 0.23$\,km\,s$^{-1}$ and
log\,$A_{\rm C} = 8.40 \pm 0.13$, for which  the mean abundance
correction is null as confirmed in Fig.\,4b.
On the other hand, the intermediate-strength lines 
remain unexplained. We previously concluded that
lines with $ {\rm log} W/\nu > -4.5$, most of which originate
in the low excitation levels, should include excess absorption 
originating in the outer molecular layers. It is, however, unclear
if this interpretation can be extended to the 
intermediate-strength lines and we return to this subject in 
Sect.\,7.2  after we examine more cases.

The results of the same analysis on other late M giants show
similar patterns as shown for  $\delta^2$ Lyr and 30g Her in
Figs.\,5a-b.  
It appears that the abundance corrections for the second overtones
(filled triangles) are often larger than those
for the first overtones (filled circles), but we assume that
such a difference may be within the margins of error of the measurements,
including the difference in the assumed continuum levels.
The case of RX Boo shown in Fig.\,5c is the most difficult case
with only 3 weak lines, but nevertheless we could find a solution
in this case. But the PEs both of $\xi_{\rm micro}$ and 
log\,$A_{\rm C}$, estimated by the method noted in Sect.\,3 are,
however,  quite large reflecting the poor quality of the data 
available for this latest M giant in our sample.

\begin{figure}
\centering
\includegraphics[width=8.0cm]{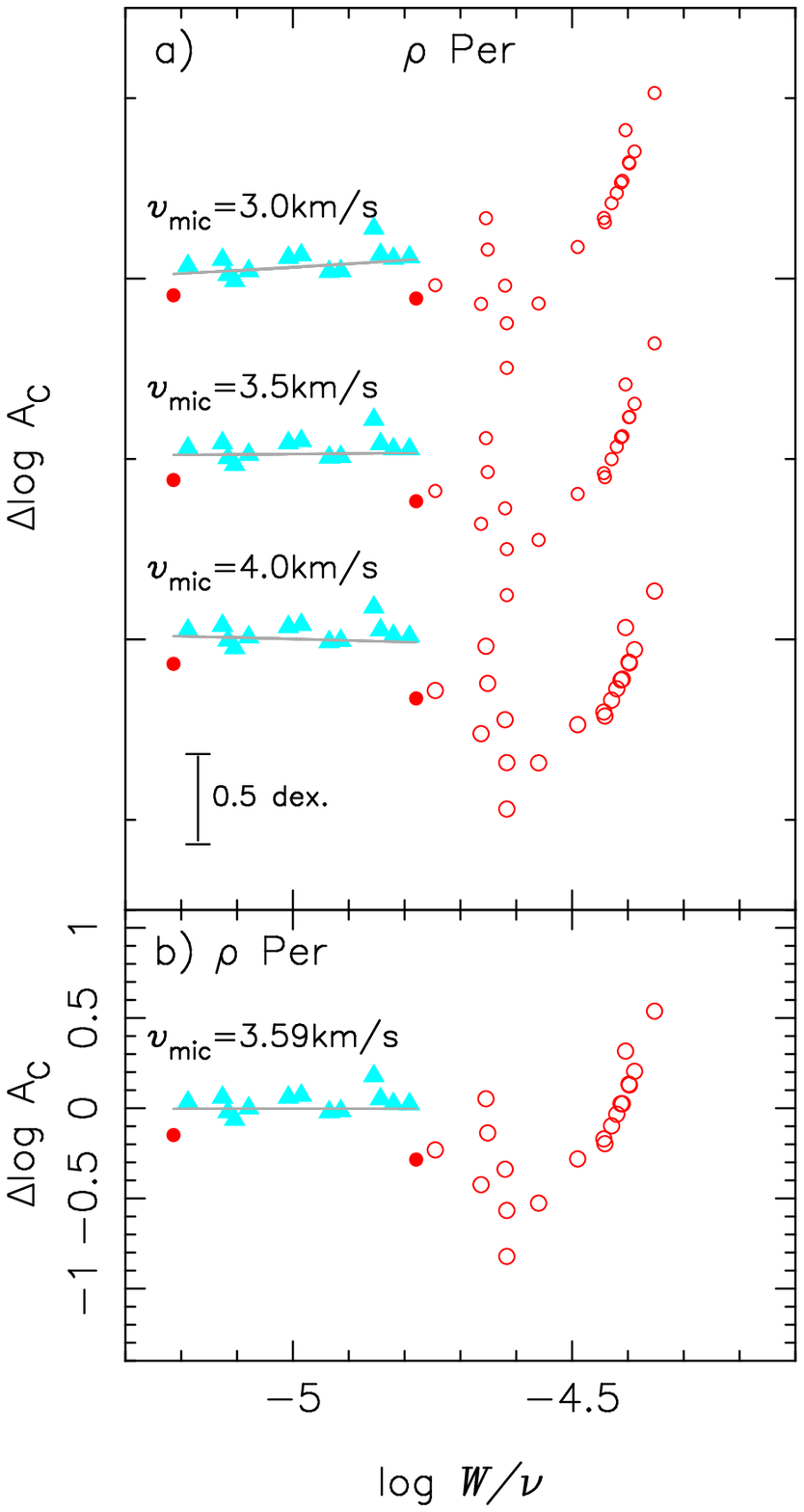}
\caption{
{\bf a)} Logarithmic abundance corrections  for the lines of the CO
first and second overtone bands  in  $\rho$ Per plotted against
the observed $ {\rm log} W/\nu $ values for assumed values of
$\xi_{\rm micro}$ = 3.0, 3.5, and 4.0 km\,s$^{-1}$ (model: b/2.0/100/3500).
{\bf b)}  Confirmation of the null logarithmic  abundance corrections  for
log\,$A_{\rm C}$ = 8.27  and $\xi_{\rm micro}$ = 3.59\,km\,s$^{-1}$, which
are the solution of the line-by-line analysis of the weak lines (filled
symbols) in a).
}
\label{Fig8.eps}
\end{figure}

\begin{figure}
\centering
\includegraphics[width=8.0cm]{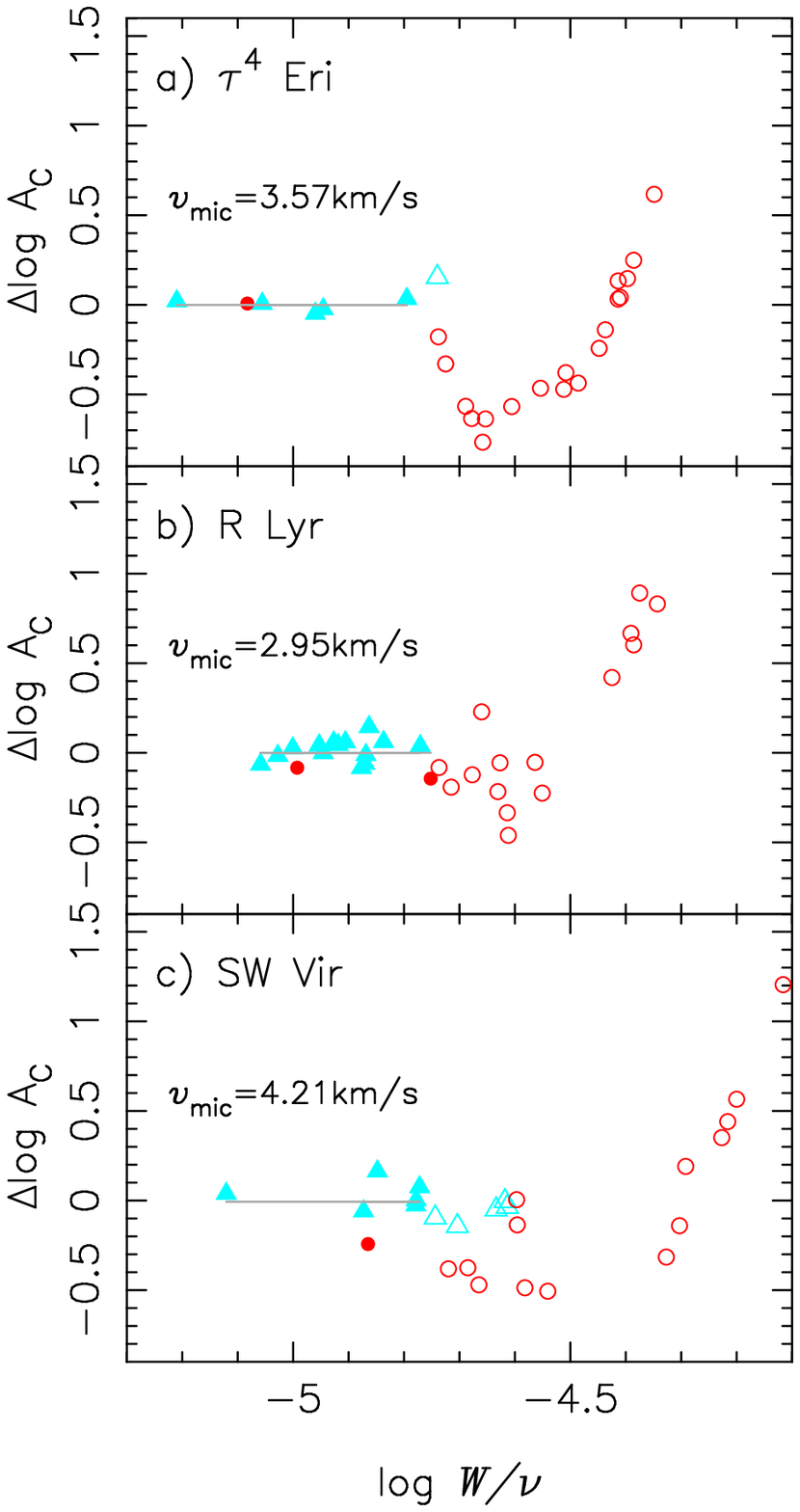}
\caption{
{\bf a)} Confirmation of the null logarithmic  abundance corrections
for log\,$A_{\rm C}$ = 8.41 and $\xi_{\rm micro}$ = 3.57\,km\,s$^{-1}$,
resulting from the line-by-line analysis of the weak lines of CO in
$\tau^4$ Eri (model: a/2.0/75/3700).
{\bf b)} The same as a) but for log\,$A_{\rm C}$ = 8.26 and
$\xi_{\rm micro}$ = 2.95\,km\,s$^{-1}$ in R Lyr (model: b/2.0/150/3300).
{\bf c)} The same as a) but for  log\,$A_{\rm C}$ = 8.26 and
$\xi_{\rm micro}$ = 4.21\,km\,s$^{-1}$  in SW Vir (model: b/2.0/250/2900).
}
\label{Fig9.eps}
\end{figure}

We then examine earlier M giants and  apply the same analysis 
to the lines of the CO first and second overtone bands of 
$\beta$ Peg, as shown in Fig.\,6a. We applied the same method of the 
line-by-line analysis to the first and second overtone bands separately 
in Fig.\,1c of Paper I  and Fig.\,2 of Paper III, respectively.
The lines of the second overtone bands were all weak lines 
and  the results were $\xi_{\rm micro} =
4.1 \pm 2.8$\,km\,s$^{-1}$ and log\,$A_{\rm C} = 8.24 \pm 0.08$ (Paper III).  
Although the lines of the first overtone bands covered
the range of $ -5.1 < {\rm log}\,W/\nu < -4.5 $, all these
lines could also be analyzed consistently, and the results were
found to be $ \xi_{\rm micro} = 3.12$\,km\,s$^{-1}$ and 
log\,$A_{\rm C} = 7.89$ (Paper I). Now, exactly the same 
lines of the first and second overtone bands are analyzed together and 
our results for $ \xi_{\rm micro}$ = 1.6, 2.0 and 2.4 km\,$^{-1}$ 
are shown in Fig.\,6a. It appears again  that no consistent solution can be 
found  for all these lines covering $ -5.3 < {\rm log}\,W/\nu < -4.4 $
\footnote{Note that our previous analysis of the CO first overtone lines 
with $ -5.1 < {\rm log} W/\nu < -4.5$  appeared to provide 
a solution (Fig.\,1c of Paper I), but it now appears to be a wrong  
solution because of the two reasons. First, as we have just found, it 
included the intermediate-strength lines to which the line-by-line 
analysis based on the classical micro-turbulent model cannot be applied.
Second, it included only few weak lines. With such 
lines of the CO first overtone bands, the line-by-line analysis 
converged to an incorrect solution, and this is an example of a 
systematic error due to an improper selection of lines, as noted 
in Sect.\,3. Such a convergence to a wrong solution could  be prevented
and a correct critical value could  be found, if a sufficient 
number of weak lines could have been included in our analysis of
Paper I.
}. 
Instead the lines are divided into two groups separated at
$ {\rm log} W/\nu \approx -4.75$, and a consistent abundance
correction can be obtained  from only the weak lines. 
The resulting $\xi_{\rm micro} = 2.04 \pm 0.07$km\,s$^{-1}$ and 
log\,$A_{\rm C} = 8.27 \pm 0.05 $ are confirmed to provide
null abundance corrections as shown in Fig.\,6b.
The resulting carbon abundance  agrees well with that of Paper III but 
differs substantially from that of Paper I.  It is now clear that the 
analysis of Paper I  included the lines of $ {\rm log} W/\nu >-4.75$,
which were inappropriate for abundance analyses.

The results of the same analysis on other early M giants (including a K
giant) show similar patterns. Some examples are shown for
$\alpha$ Tau,  $\nu$ Vir, and $\alpha$ Cet in Figs.\,7a-c. It is to be 
noted that the anomalous behavior of the lines stronger than
the critical value already appears in the K5 giant $\alpha$ Tau, 
and in all the early M giants that we have examined. The nature of the 
intermediate-strength lines, however, is quite difficult to understand.
We have previously suggested that the lines with $ {\rm log}\,W/\nu >
-4.5$ may be contaminated by the contributions of the warm molecular
layers above the photosphere, but this interpretation has so far 
been considered only for low excitation lines observed in late M giants
(Paper II). It is to be noted, however, that the  anomalous behaviors of
the intermediate-strength lines   shown on the
$\Delta$\,log\,$A$ - log\,$W/\nu$ diagrams are quite similar between
the late (Figs.\,4 \& 5) and early (Figs.\,6 \& 7) M giants.  
 
Finally, we noticed that the case of $\rho$ Per is somewhat different.
We again apply the line-by-line analysis for $\xi_{\rm micro}$ = 3.0, 3.5 
and 4.0\,km\,s$^{-1}$ as shown in Fig.\,8a.
The resulting  $\xi_{\rm micro} = 3.59 \pm 0.64$\,km\,s$^{-1}$ and 
log\,$A_{\rm C} = 8.27 \pm 0.07$ based on the weak lines 
are confirmed in Fig.\,8b.  It is to be remembered that exactly 
the same lines of the first overtone bands alone (but omitting 
lines of $ {\rm log} W/\nu > -4.5$) resulted
$\xi_{\rm micro} = 2.68 \pm 0.30$\,km\,s$^{-1}$ and log\,$ A_{\rm C} 
= 8.00 \pm 0.22 $ (see Fig.\,1e of Paper I), although it now appears that 
almost all the lines used in Paper I were the intermediate-strength lines,
which may be inappropriate
for abundance analysis. Although the weak lines 
behave as expected,  the intermediate-strength lines 
do not at all in that the abundance corrections are now mostly negative,
while they were mostly positive in the cases examined so far
(see Figs.\,4 - 7).

We found  more or less similar patterns in other objects such as
$ \tau^4 $ Eri, R Lyr, and SW Vir shown in Figs.\,9a-c.
We note that the presence of the warm molecular layers
is already known in late M giants including $ \rho $ Per, R Lyr, and SW Vir
(Paper II) and the strong low excitation lines  
in Figs.\,8 \& 9 show positive abundance corrections consistent with the
excess absorption in the outer molecular layers, even though
weaker intermediate-strength lines show negative abundance corrections.
Thus, the presence of the extra molecular layers in these
objects can still be considered, but we will return to this subject
after we also analyze OH lines (Sect.\,7.2). 

We conclude that the weak lines of CO behave quite well as expected from 
the classical micro-turbulent model described in Sect.\,3
and hence photospheric abundances can be determined with reasonable
accuracy from these lines, although the behaviors of the 
intermediate-strength lines cannot yet be understood. Thus, our 
initial purpose to determine carbon abundance could be achieved, and
we summarize the resulting carbon abundances based on the lines of
the CO first and second overtone bands for 23 objects in the 6-th column of
Table 6. We found that the results do not agree at all with those of Paper I
but agree rather well with those of Paper III except for a few cases.
We conclude that the carbon abundances of Paper I cannot be
correct because they were based on the lines that we found to be 
inappropriate for abundance analysis on a purely empirical basis.

Although  our new ${\rm log}\,A_{\rm C}$ values agree rather well
with those of paper III, we notice  for a few objects that there are 
differences as large as 0.3 dex, which have a direct effect on the 
determination of the nitrogen abundances based on CN lines (Aoki \& Tsuji
1997). Roughly speaking, if $ P_{\rm C} $ and $ P_{\rm N} $ are
determined by CO and N$_{2}$ formations, respectively,
 $$ P_{\rm CN} \propto  P_{\rm C} P_{\rm N} \propto A_{\rm C} 
A_{\rm N}^{1/2}, $$
where  $ P_{\rm C}$, $ P_{\rm N}$, and $ P_{\rm CN}$ are the partial
pressures of C, N, and CN, respectively.
Then, $ P_{\rm CN}$ remains unchanged if a change of $A_{\rm C}$ by 
$\Delta {\rm log}\,A_{\rm C}$ is compensated for by that of  
$A_{\rm N}$ by $\Delta {\rm log}\,A_{\rm N}  \approx 
-2\,\Delta {\rm log}\,A_{\rm C}$ . For a more accurate assessment of the
effect of the revised $A_{\rm C}$  on $A_{\rm N}$, 
we use the line intensity  integral  ${\it\Gamma_{\nu}(\chi)}$ 
(see the Appendix of Paper III) defined so that
$$ {\rm log}\,(W/\nu)_{\rm wk} = {\rm log}\,gf +{\it \Gamma_{\nu}(\chi)}. $$ 
 We determine a new $A_{\rm N}$, with the revised $A_{\rm C}$,
 so that it produces the same   ${\it \Gamma_{\nu}(\chi)}$ based on the
previous values of $A_{\rm C}$ (Paper III) and 
$A_{\rm N}$ (Aoki \& Tsuji 1997). It is impossible to
do this analytically, but a few trials and  errors are sufficient to
find a solution. We apply this correction in a typical line with the
PP models used previously for the determination of the N abundances. 
The resulting change in ${\rm log}\,A_{\rm N}$ 
is found  to be between -$\Delta {\rm log}\,A_{\rm C}$ and
-2\,$\Delta {\rm log}\,A_{\rm C}$, and the revised values of
${\rm log}\,A_{\rm N}$ based on the lines of CN Red System  are given
in the 14-th column of Table 6.

\begin{table*} [ht]
\caption{ Carbon and oxygen abundances and micro-turbulent velocities 
(with probable errors(PEs) in 23 red giant stars}
\vspace{-2mm}
\begin{tabular}{ l c c c c  c c c c c c c l c}
\hline \hline
\noalign{\smallskip}
   star & $ N_{\rm H}^{~~a}$ & $ N_{\rm H}^{wk}$ & $ N_{\rm K}^{~~b}$ & 
$ N_{\rm K}^{wk}$ &  log $A_{\rm C} $ & $\xi_{\rm micro}$ &
 $ N_{\rm H}^{~~c}$ & $ N_{\rm H}^{wk}$ & $ N_{\rm L}^{~~d}$ & 
$ N_{\rm L}^{wk}$ &  log $A_{\rm O} $ & ~~$\xi_{\rm micro}$ &
log $A_{\rm N}^{~~f} $  \\    
      &   &  &  &  &  & ~km\,s$^{-1}$ &  &  &  & & &~ ~km\,s$^{-1}$ &  \\ 
\noalign{\smallskip}
\hline
\noalign{\smallskip}  
  $\alpha$ Tau  & 13 & 13 & 25 & 9  & 8.38 $\pm$ 0.04 & 2.20$\pm$ 0.09
 & ...  & ...  & 28  & 25 & 8.79 $\pm$ 0.04 & 3.85 $\pm$ 0.13 & 8.05 \\
                &    &    &    &    &                  &
 & 23   & 21  & ...  & ...  & 8.76 $\pm$ 0.09  & 3.81 $\pm$ 0.94 & \\
  $\delta$ Oph  &  5 &  5 &  21 & 6 & 8.27  $\pm$ 0.08 & 2.46$\pm$ 0.16
 & 42 &  27  & ...  & ...  & 8.83 $\pm$ 0.14 & 2.84 $\pm$ 0.54 & 8.58 \\
  $\nu$ Vir   &  6 &  6 & 22  & 10  & 8.13  $\pm$ 0.08 & 2.14$\pm$ 0.15
 & 55  & 20 & ...  & ...  & 8.84 $\pm$ 0.05 & 2.67 $\pm$ 0.24 & 8.11\\
  $\alpha$ Cet  &  6 &  6 & 11 & 3  & 8.64  $\pm$ 0.05 & 1.95$\pm$ 0.08
 & 38  & 18  & ...  & ...  & 8.98 $\pm$ 0.14 & 4.25 $\pm$ 1.03 & 7.73\\
  $\sigma$ Lib  & 12 & 12 & 14  & 2  & 8.23 $\pm$ 0.03 &  2.09$\pm$ 0.13
 & 51  & 19  &...   & ...  & 8.82 $\pm$ 0.10 & 2.03 $\pm$ 0.20 & 8.15\\
  $\lambda$ Aqr  &  5 &  5 & 13 &  4 & 8.56 $\pm$ 0.11 & 1.97 $\pm$ 0.15
 & 30 & 7  & ...  & ...  & 8.92 $\pm$ 0.15 & 6.24 $\pm$ 6.01 & 7.88\\
  $\beta$ Peg & 12 & 12 & 29 &  7  & 8.27 $\pm$ 0.05 & 2.04 $\pm$ 0.07     
 & 53 & 16   & ...  & ...  & 8.77 $\pm$ 0.12 & 3.12 $\pm$ 0.72 & 8.11\\
  $\tau^{4}$ Eri &  6  & 5 & 20 &  1  & 8.41 $\pm$ 0.04 & 3.57 $\pm$ 0.46 
 & 25 & 9   & ...  & ...  & 8.87 $\pm$ 0.19 & 3.38 $\pm$ 1.32 & 7.94\\ 
  $\mu$ Gem   & 19 & 19 & 34 & 9  & 8.32 $\pm$ 0.06 & 2.00 $\pm$ 0.11 
 & 36 & 16   & ...  & ...  & 8.81 $\pm$ 0.13 & 2.58 $\pm$ 0.33 & ... \\ 
  $\delta$ Vir  & 16 & 16 & 19 &  4  & 8.50 $\pm$ 0.06 & 2.20 $\pm$ 0.11
 & 36 & 11   & ...  & ...  & 8.84 $\pm$ 0.15 & 2.47 $\pm$ 0.45 & 7.90\\
  10 Dra   &  5 &  5 & 15 &  3  & 8.43 $\pm$ 0.14 & 2.69 $\pm$ 0.26   
 & 30 & 15   & ...  & ...  & 8.92 $\pm$ 0.18 & 2.44 $\pm$ 0.41 & 8.11\\
  $\rho$ Per  & 13 & 13 &  22 & 2 & 8.27 $\pm$ 0.07 & 3.59 $\pm$ 0.64   
 & 43  & 7  & ...  & ...  & 8.87 $\pm$ 0.07 & 3.11 $\pm$ 0.98 & 8.15 \\
  BS6861   &  5  & 4 & 12 & 2 & 8.38 $\pm$ 0.13 & 4.91 $\pm$ 1.29 
 & ...  & ...  & ...  & ...  & ...  & ... & ... \\ 
  $\delta^{2}$ Lyr &  5 &  5 &  13 & 1 & 8.22 $\pm$ 0.18 & 2.86 $\pm$ 0.40 
 & 19  & ~~3$^{~e}$  & ...   & ...  & ...  & ...& 8.41 \\ 
  RR UMi   & 15 & 15 & 15 &  4 & 8.26 $\pm$ 0.12 & 2.48 $\pm$ 0.23 
 & 34 & 5   & ...  & ...  & 8.76 $\pm$ 1.05 & 2.66 $\pm$ 2.60 & 8.07 \\ 
  $\alpha$ Her  & 11 & 10 & 17 &  3  & 8.40 $\pm$ 0.13 & 2.82 $\pm$ 0.23   
 & 22  & 0  & 42 & 16  & 8.85 $\pm$ 0.05 & 2.92 $\pm$ 0.09 & 8.14 \\ 
  OP Her   & ...  & ...  & 21 &  4  & 8.27 $\pm$ 0.60 & 2.53 $\pm$ 0.72 
 & ...  & ...  & ...  & ...  & ...   & ... & ... \\
  XY Lyr   & ...  & ...  & 14 & 4 & 8.15 $\pm$ 0.15 & 2.96 $\pm$ 0.24   
 & ...  & ...  & ...  & ...  & ...  & ...  & ...\\
  R Lyr    & 14 & 14 &  17 & 2  & 8.26 $\pm$ 0.08 & 2.95 $\pm$ 0.24 
 & ...  & ...  & 52 & 31   & 8.61 $\pm$ 0.04 & 2.99 $\pm$ 0.10 & 8.06 \\
                &    &    &    &    &                  &
 & 25   & 4  & ...  & ...  & 8.62 $\pm$ 0.05 & 6.53 $\pm$ 3.20 & \\ 
  RZ Ari   & 12 & 12 &  26 & 7  & 8.10 $\pm$ 0.09 & 2.54 $\pm$ 0.15 
 & ...  & ...  & 28 & 15  & 8.57 $\pm$ 0.14 & 3.35 $\pm$ 0.28 & 8.44 \\
                &    &    &    &    &                  &
 & 52 & 21  & ...  & ...  & 8.67 $\pm$ 0.07 & 3.49 $\pm$ 0.50 &  \\
  30g Her  & 6 &  5 & 24 &  3  & 8.35 $\pm$ 0.11 & 2.61 $\pm$ 0.26   
 & 23  & 0  & 38 & 26   & 8.75 $\pm$ 0.09 & 2.49 $\pm$ 0.16 & 7.92 \\
  SW Vir   &  11 & 6  &  15 & 1 & 8.26 $\pm$ 0.20 & 4.21 $\pm$ 2.38 
 & ...  & ...  & 29 & 16  & 8.54 $\pm$ 0.11 & 1.83 $\pm$ 0.71 & 8.46 \\ 
  RX Boo   &  6 & 2  &  13 & 1 & 8.32 $\pm$ 1.62 & 3.41 $\pm$ 2.38 
 & ...  & ...  & 13 & 12  & 8.55 $\pm$ 0.10 & 2.04 $\pm$ 0.18 & 8.26\\
\noalign{\smallskip}
\hline \hline
\noalign{\smallskip}
\end{tabular}
\vspace{-2mm}
\begin{list}{}{}
\item[$^{\mathrm{a}}$]   The number of CO lines measured in the $H$ band 
region, of which the number of weak lines (log\,$W/\nu \la -4.75$) is 
given in the next column. 
\item[$^{\mathrm{b}}$] The same as $^a$ but in the $K$ band region.
\item[$^{\mathrm{c}}$] The same as $^a$ but for OH lines in the $H$ band
                        region.
\item[$^{\mathrm{d}}$] The same as $^a$ but for OH lines in the $L$ band
			 region.
\item[$^{\mathrm{e}}$] with only 3 weak lines of similar intensities, the
		micro-turbulent velocity cannot be fixed by our
			 line-by-line analysis.
\item[$^{\mathrm{f}}$] N abundances based on CN 
(Aoki \& Tsuji 1997) corrected for the effect of the revised C abundances. 
\end{list}
\vspace{-2mm}
\end{table*}

\subsection{Uncertainties}

 So far, we have only considered internal errors (PEs) in our results
shown in Table 6, but the PEs are useful indicators of the reliability 
of the results, as noted in Sect.\,3. For example, inspection of Table 6 
reveals that the PEs for  $\xi_{\rm micro}$ are larger than
1\,km\,s$^{-1}$ for BS6861, SW Vir, and RX Boo, and this may be because
the numbers of weak lines available in these cases are only a few. 
A large systematic error due to an improper selection of lines in 
Paper I is corrected with the recognition of the possibly true
critical value of log\,$W/\nu = -4.75$ in this paper. 
Other possible uncertainties are due to errors
in the input parameters such as $T_\mathrm{eff}$, $ M $, $ R $,  
and to our modeling of the photospheres.   
We previously examined the effect of uncertainties in 
$T_\mathrm{eff}$  and ${\rm log}\,g $ within the 
framework of  PP models. The results (Fig.\,14 of Paper III)
indicated that the uncertainties are $ \Delta\,{\rm log}\,A_{\rm C}
 \la 0.1 $ for $ \Delta T_\mathrm{eff} = 100$\,K and 
$\Delta\,{\rm log}\,A_{\rm C} \approx 0.2 $ for $\Delta {\rm log}\,g =
0.5 $. 

The use of SS models, however, introduces
another problem: Inspection of Table 5 reveals that
the photospheric extensions and hence surface temperatures show
considerable differences between models of one and two solar masses   
especially at low $T_\mathrm{eff}$.  In  the model photospheres of 
$T_\mathrm{eff} = 3000$\,K, for example, the photospheric extension 
increases from 0.20\,$R$ to 0.51\,$R$ and the surface temperature decreases
from 1342\,K to 1079\,K if mass is reduced from  $ 2.0\,M_{\sun} $ to
$ 1.0\,M_{\sun} $ (Table 5). We  assume 2.0\,$M_{\sun}$ in our
analysis in this paper throughout, but mass is the most uncertain parameter.  
For this reason, we examined the effect of mass for a typical case
of $\alpha$ Her as an example. We obtained the results shown in Table 6 
(${\rm log}\,A_{\rm C} = 8.40 $ and $ \xi_{\rm micro} =2.82$\,km\,s$^{-1}$) 
with a model of $ 2.0\,M_{\sun} $ (b/2.0/150/3300).  We repeat 
the same analysis with a model of $ M = 1.0\,M_{\sun}$ ( b/1.0/150/3300).
The result is ${\rm log}\,A_{\rm C} = 8.29 \pm 0.13 $ and
$\xi_{\rm micro} = 3.05 \pm 0.24$\,km\,s$^{-1}$. The difference in
derived carbon abundance is 0.11 dex and turbulent velocities differ
by less than 10\,\%. Thus, the conclusion obtained with PP
models remains almost unchanged if we use SS models.  
The reason for this may be due to the use of only
the weak lines, which are formed rather deep in the photosphere. 
Also CO is quite stable at low temperatures because of its high 
dissociation energy; its abundance may be hardly affected by 
changes of the  physical conditions. This is in 
marked contrast to the case of the Sun in which CO is
just forming at temperatures considerably higher than in M giants
and its abundance is highly sensitive to temperature.

\subsection{Comparisons with other author's results}
We compare our carbon abundances with those by Smith \& Lambert
(1990) for objects analyzed in both analyses presented in Table 8.  
 Some significant  differences are found for
$\nu$ Vir,  $\beta$ Peg,  and $\rho$ Per, which show
differences of 0.38, 0.15, and 0.19 dex, respectively.
The agreements for other objects, however, are generally within
the internal errors of both  analyses.
Unfortunately, we are unaware of other high resolution studies 
of photospheric abundances of cool giant stars. 

We note, however, an interesting intermediate-resolution analysis
for a large sample of 70 M giant stars (Lazaro et al. 1991). Their 
result implied  rather low carbon abundances consistent with 
our previous result of Paper I, which we just discarded. Since their
intermediate-resolution analysis necessarily included the 
intermediate-strength lines as we also did in our previous analysis 
in Paper I, their result demonstrates that the low and high
resolution analyses provide similar results if applied to the same 
types of lines.  Certainly, intermediate-resolution analysis should be
useful for  applications to faint objects in remote stellar systems.
However, the problem of how to treat the stronger lines that have
not been well modeled by the classical line formation theory
must be settled before lower resolution spectra can be used
successfully in abundance analyses.

Another interesting attempt of medium resolution spectroscopy
was reported by Decin et al.(2003), who analyzed infrared spectra
observed with ISO SWS.
They carefully compared the observed and synthetic spectra and 
estimated basic stellar parameters, including the CNO and their 
isotopic abundances. For synthetic spectra of medium resolution, 
it may again be difficult to avoid the use the intermediate-strength
and strong lines inappropriate for abundance analysis.
Nevertheless the resulting carbon abundances agree rather well 
with the other results except for $\alpha$ Cet, for which our 
result shows unusually high C abundance.

\begin{figure}
\centering
\includegraphics[width=8.0cm]{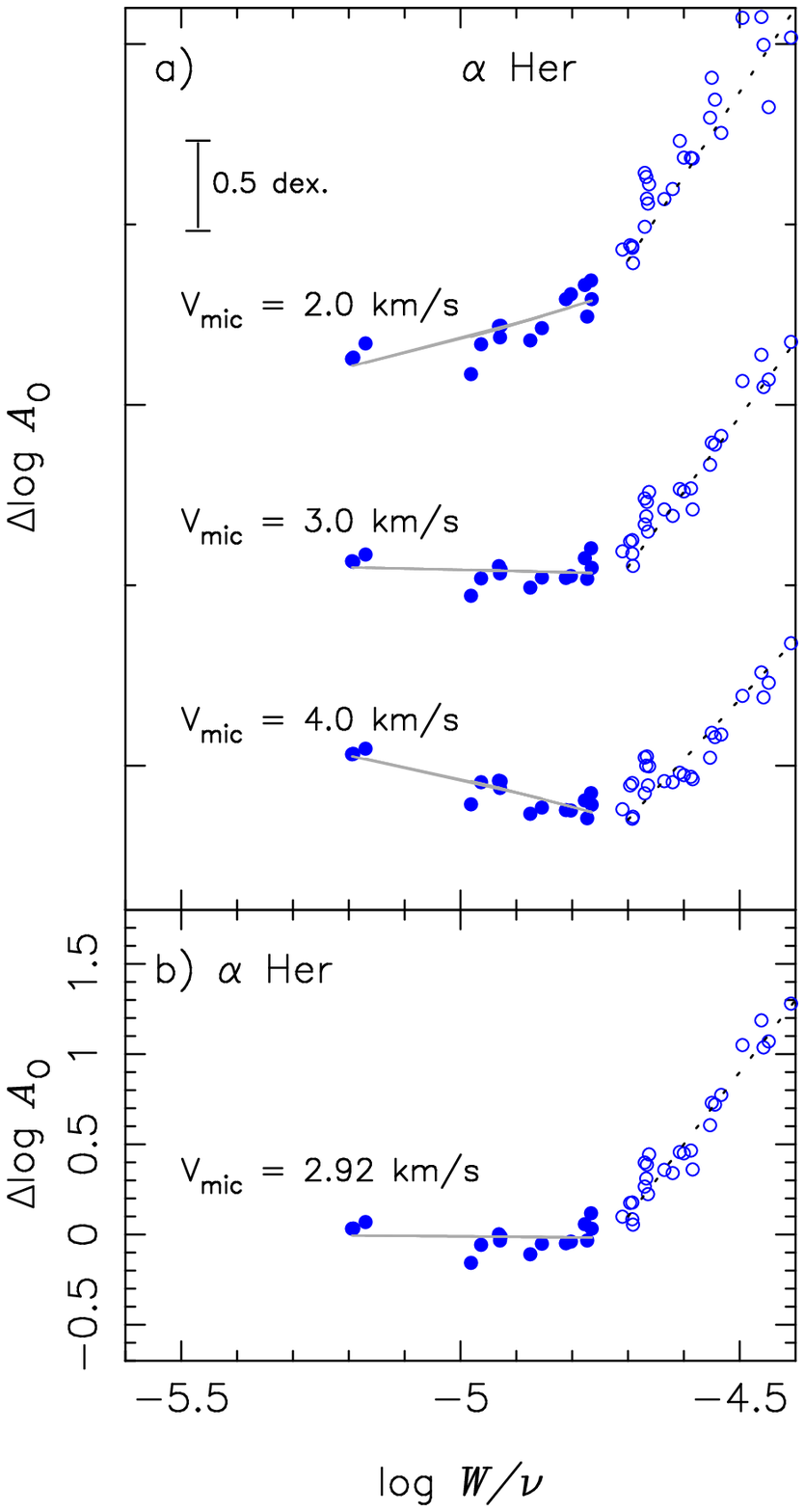}
\caption{
{\bf a)} Logarithmic abundance corrections  for the lines of OH
fundamental bands in $\alpha$ Her plotted against the observed
$ {\rm log}\,W/\nu $ values for assumed values of $\xi_{\rm micro}$ =
2.0, 3.0, and 4.0 km\,s$^{-1}$ (model: b/2.0/150/3300).
{\bf b)} Confirmation of the null logarithmic  abundance corrections  for
log\,$A_{\rm O}$ = 8.85  and $\xi_{\rm micro}$ = 2.92 km\,s$^{-1}$, which
are the solution of the line-by-line analysis of the weak lines (filled
circles) in a).
}
\label{Fig10.eps}
\end{figure}

\begin{figure}
\centering
\includegraphics[width=8.0cm]{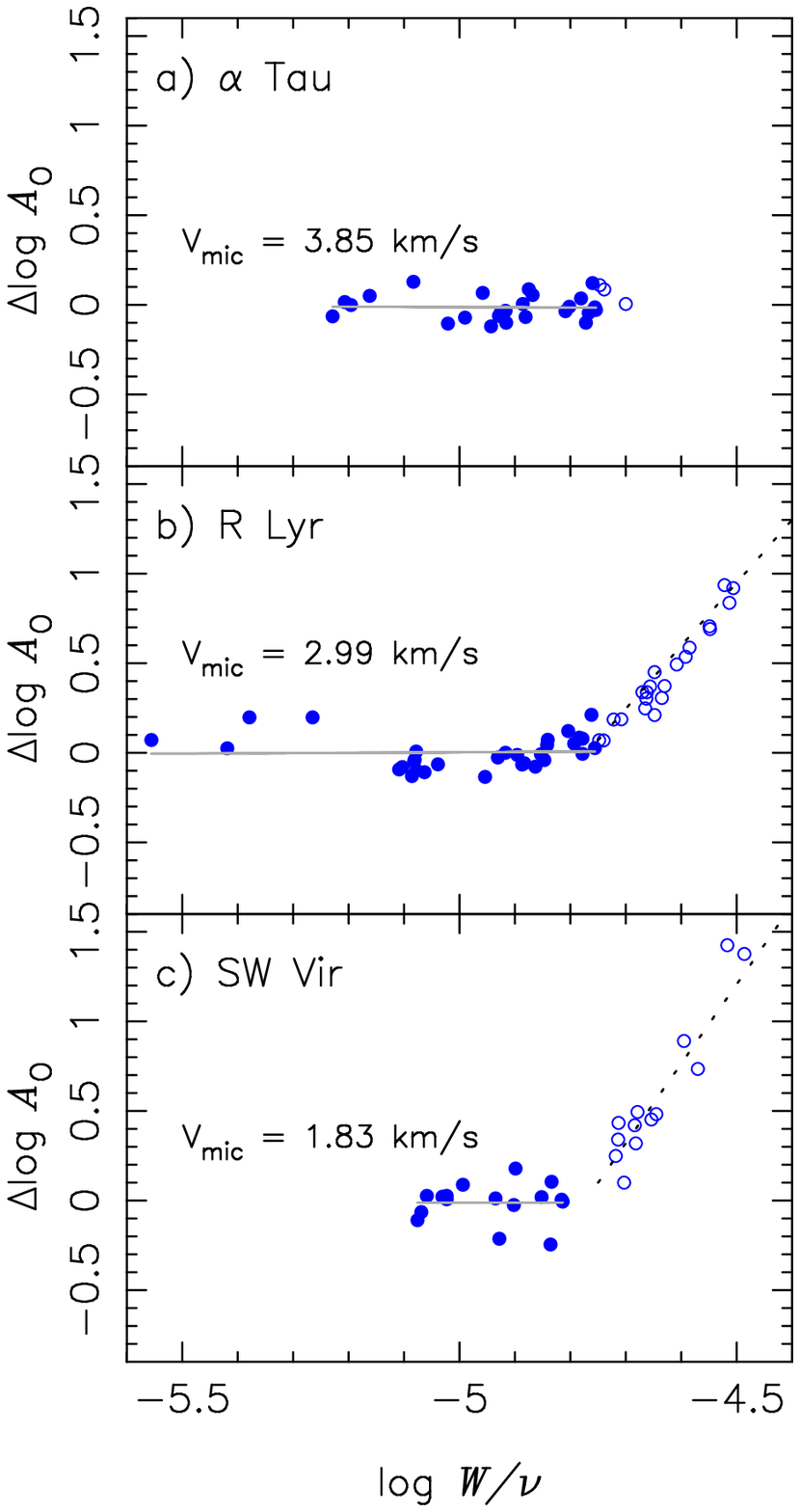}
\caption{
{\bf a)} Confirmation of the null logarithmic  abundance corrections
for log\,$A_{\rm O}$ = 8.79  and $\xi_{\rm micro}$ = 3.85 km\,s$^{-1}$,
resulting from the line-by-line analysis of the weak lines of OH in
$\alpha$ Tau (model: a/2.0/50/3900).
{\bf b)} The same as a) but for log\,$A_{\rm O}$ = 8.61  and
$\xi_{\rm micro}$ = 2.99 km\,s$^{-1}$ in R Lyr (model: b/2.0/150/3300).
{\bf c)} The same as a) but for log\,$A_{\rm O}$ = 8.54  and
$\xi_{\rm micro}$ = 1.83 km\,s$^{-1}$ in SW Vir (model: b/2.0/250/2900).
}
\label{Fig11.eps}
\end{figure}

\section{The case of OH}

Hydroxyl radical OH shows rich spectra due to ro-vibrational
transitions in the infrared region. The fundamental and first
overtone bands are observed well in the  $L$  and $H$ band regions,
respectively, in cool oxygen-rich stars. 
Spectroscopic data such as the line positions and
intensities of OH are taken from the GEISA
databank (Jacquinet-Husson et al. 1999), which 
 covers well the high excitation transitions of OH observed in stellar
spectra.

\begin{figure}
\centering
\includegraphics[width=8.0cm]{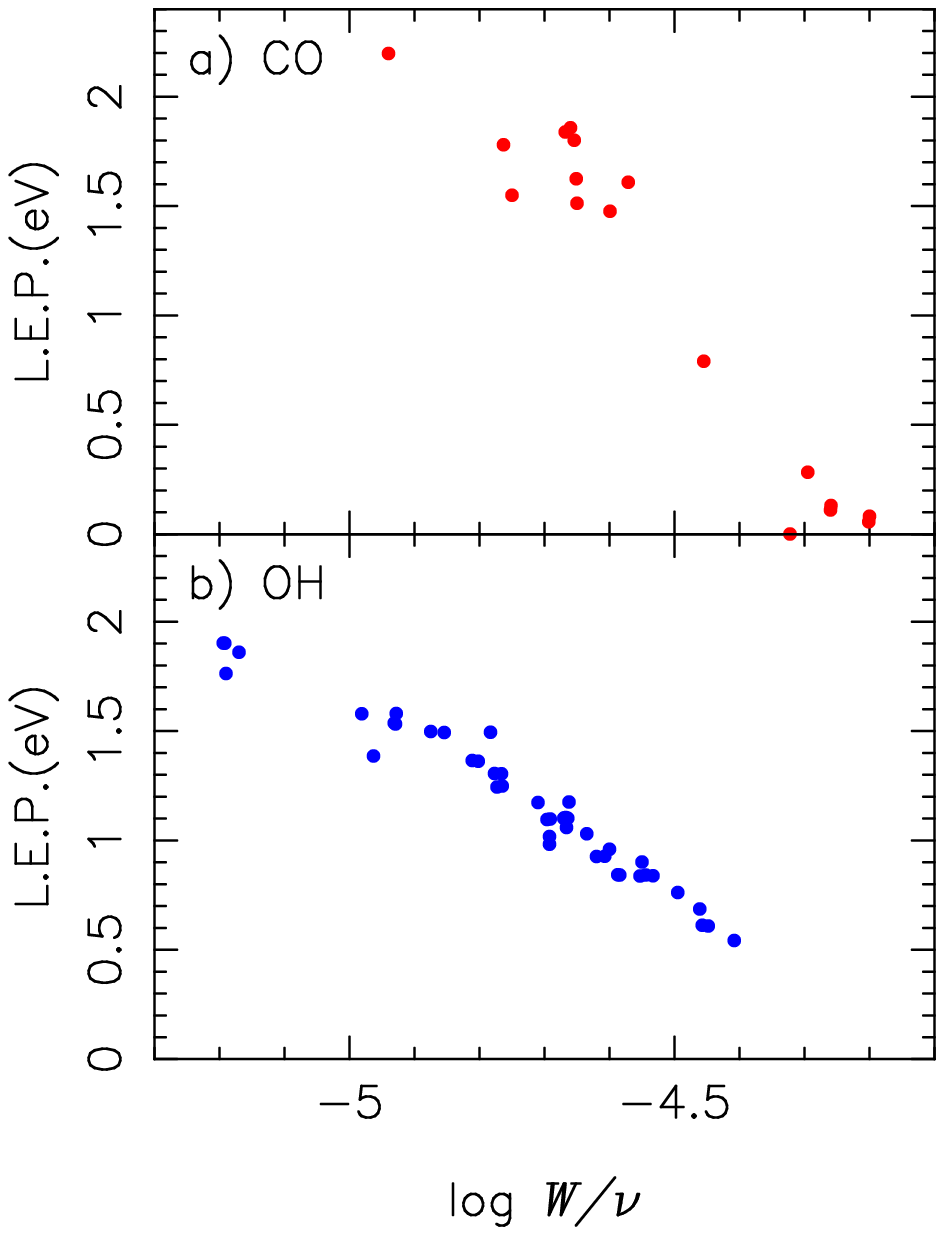}
\caption{  Lower excitation potentials (LEPs) of the lines
plotted  against  $ {\rm log}\,W/\nu$ values.
{\bf a)} CO first overtone bands observed  in $\alpha$  Her.
{\bf b)} OH fundamental bands observed in $\alpha$  Her.
}
\label{Fig12.eps}
\end{figure}

\subsection{The OH fundamental bands}

 First, we applied the same method as used for CO to the lines of the
OH  fundamentals.  We include  the carbon abundance determined in 
the previous section as an input abundance for each star. 
Since OH formation depends on $A_{\rm O} - A_{\rm C}$
rather than on $A_{\rm O}$, determination of the carbon abundance is
a prerequisite for OH analysis to determine the oxygen abundance.
An example is shown for $\alpha$ Her 
 in Fig.\,10a, in which the results of the  analysis 
with $\xi_{\rm micro}$ = 2.0, 3.0, and 4.0 km\,s$^{-1}$
are shown. We again encountered the same difficulty as for the case
of the CO first overtone bands in that the stronger lines do not follow
the pattern expected from the weaker lines. Furthermore, it appears very
clearly that  these weaker and stronger lines are divided at log\,$W/\nu
\approx -4.75$ and not at log\,$W/\nu \approx -4.5$ suggested from our
previous analysis of the CO first overtone bands alone (Paper I). 
In fact, we first  noticed that log\,$W/\nu
\approx -4.75$ should be an important value for dividing the lines
 into two groups by this analysis of the OH fundamental bands.
With this result in mind, we also confirmed that this value is similarly
important for the CO lines, but clearly seen only if the first and second
overtone bands are analyzed together (Sect.\,4.2).

As for CO, the weaker lines show the behavior expected from
the classical micro-turbulent model  noted in Sect.\,3.  
However, the stronger lines, which should suffer heavier saturation
effect, do not show the behavior expected 
for the saturated lines. Thus, we apply the
same grouping of lines as defined for CO (see footnote 5) to OH. 
From the weak lines, we found that the abundance
corrections from different lines show a consistent value for 
$\xi_{\rm micro}$ = 2.92 km\,s$^{-1}$ as shown in Fig.\,10b.  
We found again that it is impossible to smooth out the
reflection at log\,$W/\nu \approx -4.75$ by any choice of
the micro-turbulent velocity, and thus the critical value of log\,$W/\nu
\approx -4.75$ can be deemed as well defined empirically.   
This result is consistent with Smith \& Lambert (1990), who found that 
OH lines of 1-0 and 2-1 bands (mostly stronger than the lines of 3-2, 
4-3, 5-4 bands) provide unreasonably large values of the oxygen abundance 
and cannot be used for abundance determinations.    

It is unclear if the same problem exists for the OH lines of K giant 
$\alpha$ Tau,  since almost all lines measured are the weak lines.  
We recall that the anomalous behavior 
of the intermediate-strength lines of CO was already noticed in the 
K giant $\alpha$ Tau (Sect.\,4.2).
We are able to use almost all the  OH lines measured in $\alpha$ Tau in
the  abundance analysis simply because they are the  
weak lines and the result 
is shown in Fig.\,11a. However, all the other late M giants show similar
pattern as for $\alpha$ Her. As examples, the results 
for R Lyr and SW Vir are shown in Figs.\,11b and 11c, 
respectively (30g Her, RZ Ari, and RX Boo also show similar patterns). 

We assume that the photospheric abundances of oxygen could be
obtained exclusively from the weak lines, 
 and summarize the resulting oxygen abundances and turbulent velocities 
obtained from the OH fundamentals in Table 6 for 7 objects for which 
the $L$ band spectra are available (those objects with
non-zero $N_{L}^{wk}$ in the 11-th column).

Again, a problem is why the intermediate-strength lines cannot be used 
for abundance 
analysis. We suggested that some contributions from the outer molecular 
layers will be disturbing the most strong lines in the case of CO. To 
examine if this may also be the case for OH, we plotted the lower 
excitation potentials (LEPs) of all the lines in Fig.\,10a against 
log\,$W/\nu$ in Fig.\,12b. For comparison, we made a similar plot in 
Fig.\,12a for CO lines of $\alpha$ Her analyzed in Fig.\,3.
Unlike the case of the strong CO lines originating from the levels
of very low LEP less than 0.5 eV,  the intermediate-strength lines of OH 
are not necessarily originating from the very low LEP, but rather 
from the levels with LEP $\approx$ 0.5-1.5 eV. However, the 
intermediate-strength lines of CO 
are also not originating from
the levels with very low excitation levels, and the
intermediate-strength lines  of CO and OH 
may have the same origin as will be discussed in Sect.\,7.2.
 
\subsection{The OH first overtone bands}

Unfortunately, we have only 7 spectra of the $L$ band region, and
we analyze the OH first overtone bands observed in the $H$ band region
to extend our sample. Unlike the most molecules that the overtone
bands are weaker than the fundamental bands, the case of OH is
rather special in that the first overtone bands  are as strong as
the fundamental bands or even stronger. As a result, it is more difficult
to measure weak lines in the first overtone than in the fundamental
bands especially in cooler objects, and we cannot analyze
the first overtones of OH in some cool objects.   
Fortunately, most of these objects are those for which we have the $L$ band
spectra and oxygen abundances are already determined from the OH 
lines of the fundamental bands (Sect.\,5.1).

We apply the same method as used for the fundamental bands and an example
is shown in Fig.\,13a for $\rho$ Per. For this case of the 
overtone bands, compared to the case of the fundamental bands
discussed in Sect.\,5.1, it is shown more clearly that the OH lines are 
divided into two groups that behave quite differently.   
Even though abundances can be determined rather well from the weak
lines, it is hopeless to determine abundances from the
intermediate-strength lines.
It is also interesting to note that the abundance corrections
for the intermediate-strength lines show a similar behavior to those 
for the intermediate-strength  lines of CO in $\rho$ Per (Figs.\,8a-b).

In the K giant $\alpha$ Tau, most lines measured  are weak lines  
as in the fundamental bands, and we use most of these lines for our 
abundance determination (Fig.\,14a).  In the M3 giant $\sigma$ Lib, 
the abundance corrections for the intermediate-strength lines show 
large positive deviations compared with those for the weak lines, 
for which the abundance corrections converge to null for log\,$A_{\rm C}$ =
8.82  and $\xi_{\rm micro}$ = 2.03 km\,s$^{-1}$ (Fig.\,14b).
The M5 giant R Lyr shows similar pattern as for $\rho$ Per (Fig.\,14c).
The resulting  micro-turbulent velocity is unreasonably large, but
the large PE implies that such a result cannot be realistic.
This is because  small errors in the measured equivalent widths will 
give serious effect on the determination of $\xi_{\rm micro}$ value 
when only a few weak lines are available, and this is a difficulty 
of our line-by-line analysis.  Nevertheless, the abundance
can be determined rather well because of the use of the weak lines,
which depend little on  $\xi_{\rm micro}$ value.
The difficulty of measuring weak lines in the first overtones
of OH is more serious for some late M giants including 
$\delta^2$ Lyr, 30g Her, $\alpha$ Her, SW Vir, RX Boo, and we had to give 
up to analyze the $H$ band spectra of these stars.

We confirm both in the CO and OH lines that the behaviors of the 
intermediate-strength lines are quite complicated and 
show different patterns for different objects. For example, the 
intermediate-strength lines, both of OH and CO, do not necessarily 
show positive abundance corrections compared with those for the weak lines
in $\rho$ Per and R Lyr.
It is interesting to note that CO and OH lines show similar patterns
on the  $\Delta$\,log\,$A$ - log\,$W/\nu$ diagrams if  these OH and CO lines 
are measured from the same spectrum as are the cases of 
$\rho$ Per and R Lyr
\footnote{While the intermediate-strength lines of CO in R Lyr
(Fig.\,9b) and SW Vir (Fig.\,9c) show 
negative abundance corrections, the intermediate-strength lines of OH 
fundamentals in  R Lyr (Fig.\,11b) and SW Vir (Fig.\,11c) do not. But OH
fundamental bands were not observed simultaneously with CO bands
(see Table 2) and it is possible that line formations took place at different
conditions in these variable stars.}. 

We summarize the resulting oxygen abundances based on the weak lines 
of the OH first overtone bands for 15 objects in 
Table 6 (those with non-zero $N_{H}^{\rm wk}$ in the 9-th column). 
The oxygen abundances could be determined
independently from the fundamental and first overtone bands
for three objects, $\alpha$ Tau, R Lyr, and RZ Ari. The resulting
oxygen abundances, compared directly in the 12-th column of Table 6,
agree well within the internal errors of both the analyses. 
The micro-turbulent velocities
also agree well except for R Lyr, for which the reason for the 
difficulty was discussed already (see Fig.\,14c).  
The micro-turbulent velocities of $\alpha$ Tau based on the fundamental
and overtone bands agree quite well ($\xi_{\rm micro} \approx 
3.8$\,km\,s$^{-1}$), but disagree with the result obtained from the CO 
lines ($\xi_{\rm micro} \approx 2.2$\,km\,s$^{-1}$). This result
reminds us a question what the micro-turbulent velocity means.

\begin{figure}
\centering
\includegraphics[width=8.0cm]{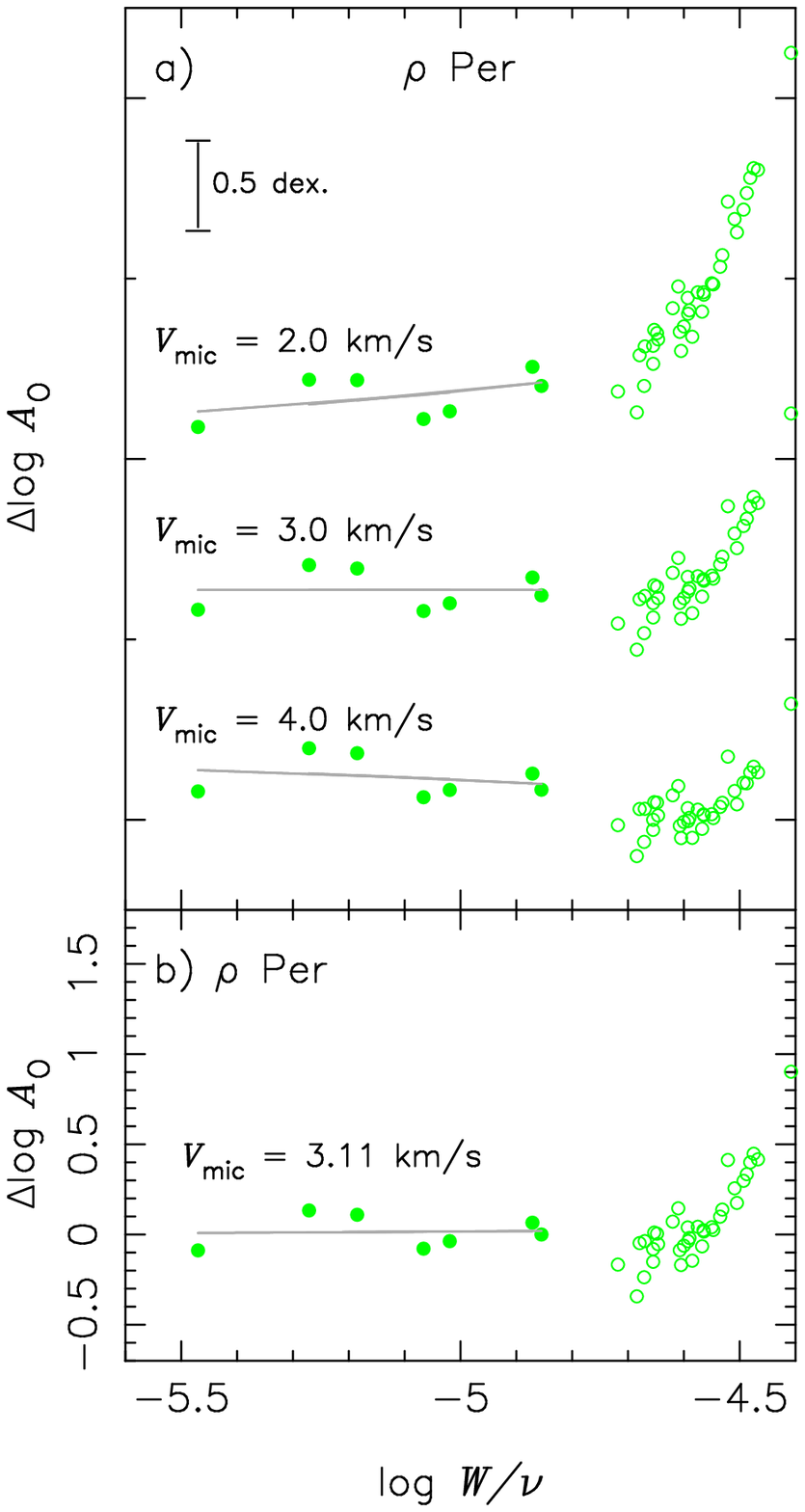}
\caption{
{\bf a)} Logarithmic abundance corrections  for the lines of the OH
first overtone  bands in $\rho$ Per plotted against the observed
$ {\rm log}\,W/\nu $ values for assumed values of $\xi_{\rm micro}$ =
2.0, 3.0, and 4.0 km\,s$^{-1}$ (model: b/2.0/100/3500).
{\bf b)} Confirmation of the null logarithmic  abundance corrections  for
log\,$A_{\rm O}$ = 8.87  and $\xi_{\rm micro}$ = 3.11 km\,s$^{-1}$, which
are the solution of the line-by-line analysis of the weak lines
(filled circles) in a).
}
\label{Fig13.eps}
\end{figure}

\begin{figure}
\centering
\includegraphics[width=8.0cm]{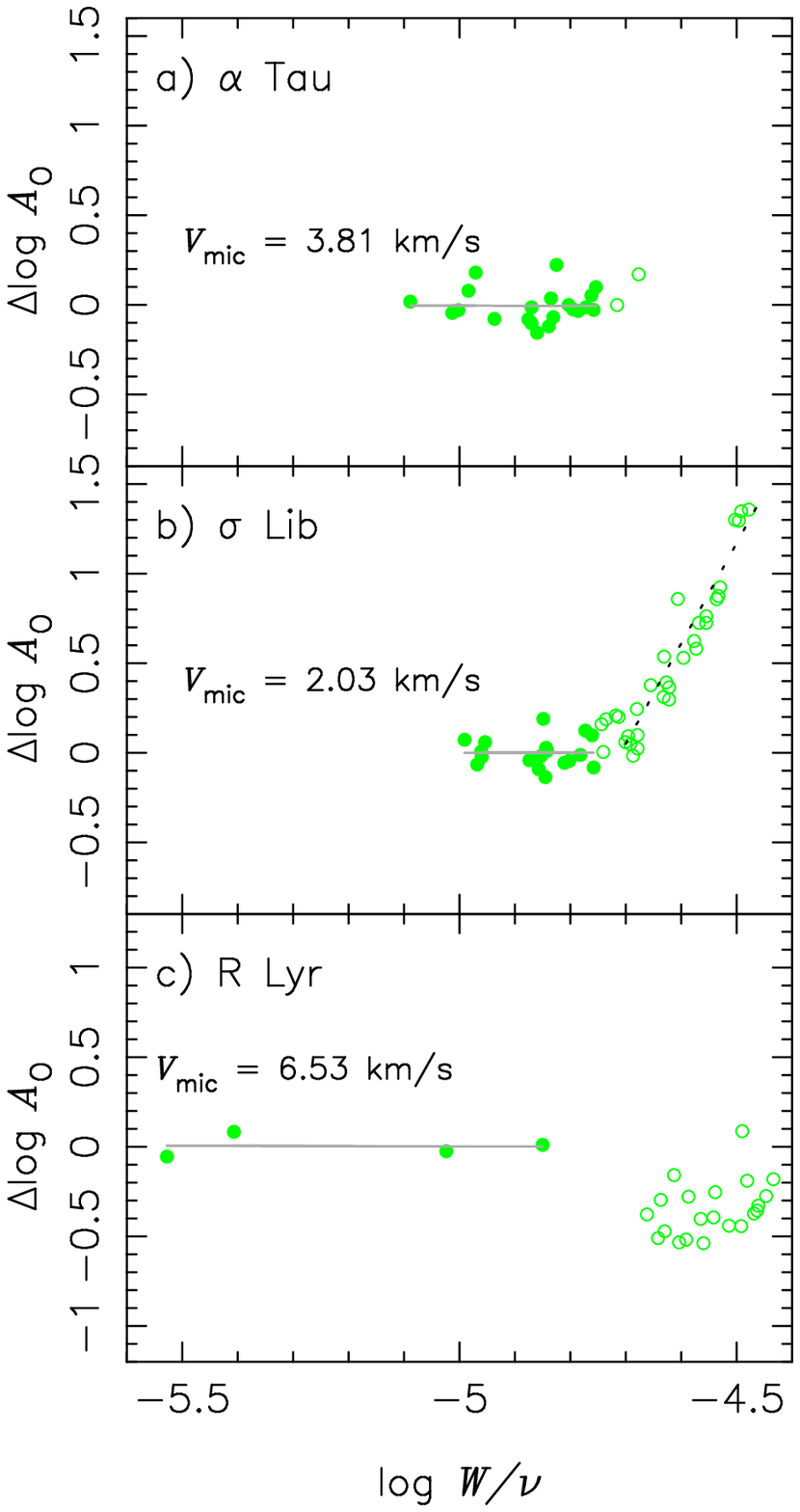}
\caption{
{\bf a)} Confirmation of the null logarithmic  abundance corrections
for log\,$A_{\rm O}$ = 8.76 and $\xi_{\rm micro}$ = 3.81 km\,s$^{-1}$,
resulting from the line-by-line analysis of the weak lines of OH in
$\alpha$ Tau (model: a/2.0/50/3900).
{\bf b)} The same as a) but for log\,$A_{\rm O}$ = 8.82  and
$\xi_{\rm micro}$ = 2.03 km\,s$^{-1}$ in $\sigma$ Lib
(model: a/2.0/100/3600).
{\bf c)} The same as a) but for log\,$A_{\rm O}$ = 8.62  and
$\xi_{\rm micro}$ = 6.53 km$^{-1}$ in R Lyr (model: b/2.0/150/3300).
}
\label{Fig14.eps}
\end{figure}

\subsection{Uncertainties and comparison with other authors}

We notice in Table 6 that the values of micro-turbulent velocity
are sometimes unexpectedly large both for CO and OH, and that
most cases of $ \xi_{\rm micro} \ga 4$\,km\,s$^{-1}$ are 
found when the numbers of weak lines used are below 10. 
The  $ \xi_{\rm micro} $ value by our analysis based on the
method outlined in Sect.\,3 is
susceptible to  small errors in the measured {\it EW}s especially if only
 a small number of lines can be used. The poor quality of the
data is reflected in the large PE of the derived $ \xi_{\rm micro} $ 
in Table 6, and the resulting $ \xi_{\rm micro} $ value cannot be  
reliable for such a case. For example, $ \xi_{\rm micro} $ values
as large as 6\,km\,s$^{-1}$  for $\lambda$ Aqr and R Lyr are almost
useless because of the very large PEs.  
  
  Unlike CO, OH may be more susceptible to a change in the
physical condition. To compare with the case of CO (Sect.\,4.3),
we again examined the  case of the cool giant  $\alpha $ Her.
We  repeat the same analysis of the OH lines of  $\alpha $ Her
 done with a model of $ M = 2.0 \,M_{\sun}$ (b/2.0/150/3300) giving 
the results shown in Table 6 (${\rm log}\,A_{\rm O} 
= 8.85 $ and $ \xi_{\rm micro} = 2.92$\,km\,s$^{-1}$). 
Now,  with a model of $ M = 1.0 \,M_{\sun}$ (b/1.0/150/3300), the 
results are ${\rm log}\,A_{\rm O} = 8.91 \pm 0.13 $ and
$\xi_{\rm micro} = 3.30 \pm 0.08$\,km\,s$^{-1}$. 
The differences are  again unexpectedly small.
The reason for this may again be due to the use of only
the weak lines, which are formed rather deep in the photosphere. 
   
Finally, we again compare our results with those by Smith \& 
Lambert\,(1990) and by Decin et al.\,(2003) in Table 8.  For most 
cases of oxygen abundances, the results agree 
reasonably well within the error bars of these analyses.   
It is interesting that the agreements are reasonably good  despite 
possible differences in the model photospheres used.

\section{Carbon and oxygen isotopic ratios}

\subsection{Line-by-line analysis}

We applied the line-by-line analysis to the  
isolated lines of $^{12}$C$^{16}$O, $^{13}$C$^{16}$O, and
$^{12}$C$^{17}$O. We use only the weak lines but include  
$^{12}$C$^{16}$O lines of the first and second overtone bands together. 
The  $\xi_{\rm micro}$ values determined from the $^{12}$C$^{16}$O 
lines of the first and second overtone bands in Sect.\,4.2 are used to 
determine abundances of $^{13}$C$^{16}$O and $^{12}$C$^{17}$O as well.
Given a $\xi_{\rm micro}$ value, abundance can be 
determined directly from each line. Some examples of the results
are shown in Fig.\,15a-d for 10 Dra, RZ Ari, $\delta$ Oph,  and $\mu$ Gem.
Unfortunately, the number of lines of the isotopic species measured is
rather small and  the weak lines could not be measured at all for 
some cases especially for $^{12}$C$^{17}$O.

    For example, the lines measured for 10 Dra are   $^{12}$C$^{17}$O
(2, 0) R\,25, R\,27, and R\,32 on the spectrum shown in Fig.\,16a. This
is a favorable case in that three $^{12}$C$^{17}$O lines could be measured,
and these lines are used for the analysis of Fig.\,15a.
In the case of RZ Ari shown in Fig.\,16b, the $^{12}$C$^{17}$O lines
are weaker and we could measure only two lines (R\,25 and R\,32) for
the line-by-line analysis of Fig\,15b. The $^{12}$C$^{17}$O lines 
are still weaker in $\lambda$ Aqr and RR UMi shown in Figs\,.16c and 16d,
respectively, and no line could be measured in these cases.
The difficulty of measuring weak lines of $^{13}$C$^{16}$O is more or
less the same, although at least one line could be measured in most
objects except for RX Boo.

The $^{12}$C/$^{13}$C and $^{16}$O/$^{17}$O  ratios based on the 
resulting $^{12}$C$^{16}$O, $^{13}$C$^{16}$O, and $^{12}$C$^{17}$O 
abundances are given 4- and 5-th columns, respectively, of Table 7
with their probable errors (PEs). 
The resulting $^{12}$C/$^{13}$C ratios are mostly larger
compared with our previous result (Tsuji 2007), and this is
because $^{12}$C$^{16}$O abundances are increased due
to the use of the lines of the first and second overtones
together.

\begin{figure*}
\centering
\includegraphics[width=13.5cm]{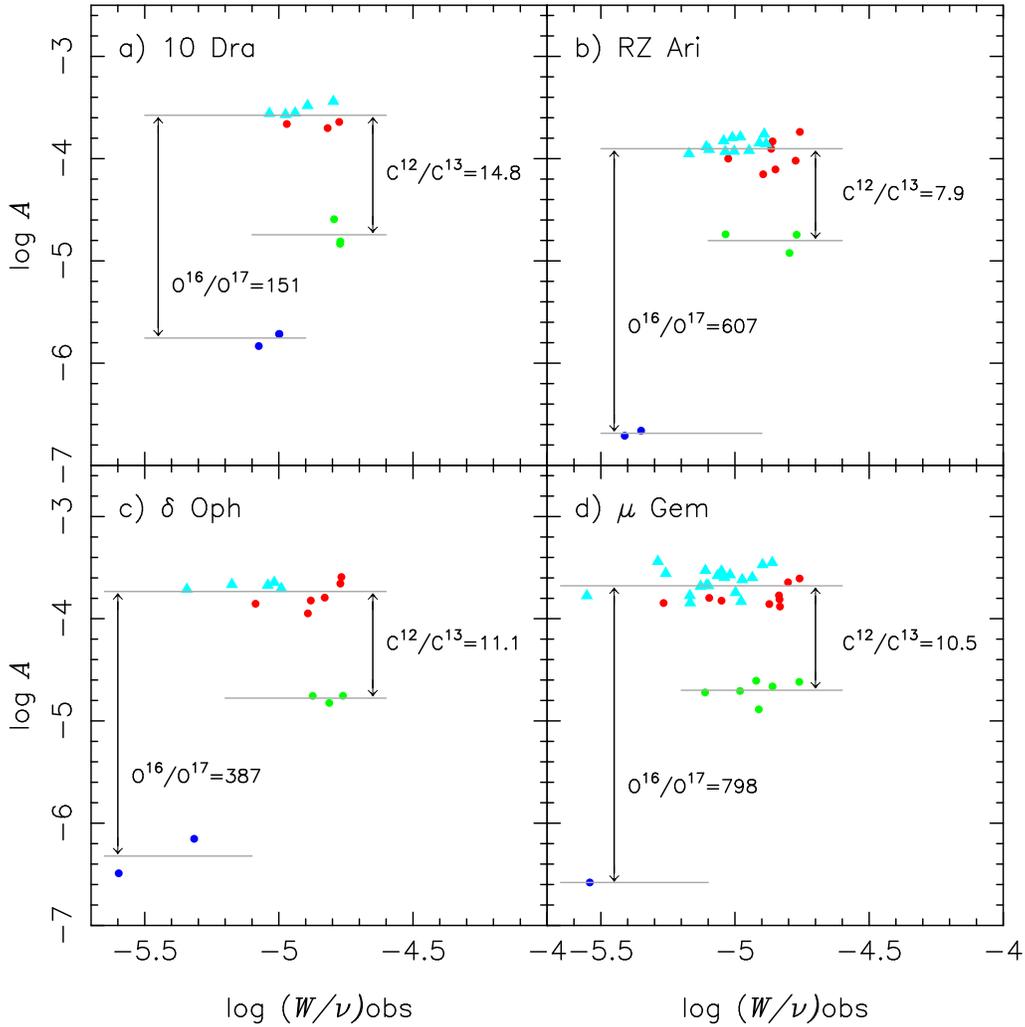}
\caption{ 
The  ordinate scale shows abundances of $^{12}$C$^{16}$O
( first and second overtones by filled circles and
triangles, respectively), $^{13}$C$^{16}$O (filled circles), and 
$^{12}$C$^{17}$O (filled circles)  
derived from the observed values of log\,$W/\nu$  in the abscissa.
{\bf a)} 10 Dra. Note that there are three lines of $^{13}$C$^{16}$O and 
$^{12}$C$^{17}$O, of which two are overlapping in the plot. 
{\bf b)} RZ Ari.  {\bf c)} $\delta$ Oph.  {\bf d)} $\mu$ Gem.
}
\label{fig15.eps}
\end{figure*}

\begin{table} 
\centering
\hspace{-20mm}
\caption{Isotopic ratios (with PEs) in 23 red giant stars }
\vspace{-2mm}
\begin{tabular}{ r c c  l l }
\hline \hline
\noalign{\smallskip}
   Obj. &  $N_{\rm wk}$(13)$^a$  &  $N_{\rm wk}$(17)$^b$ 
&  $^{12}$C/$^{13}$C  & $^{16}$O/$^{17}$O  \\ 
\noalign{\smallskip}
\hline
\noalign{\smallskip}
$\alpha$ Tau & 2  & 0 &   10.6 $\pm$ 1.0  &  $>$1000    \\
$\delta$ Oph & 3  & 2 &   11.1 $\pm$ 0.9 & ~~~387 $\pm$ 68   \\
$\nu$  Vir   & 5  & 0 &   ~8.7 $\pm$ 1.3 &   $>$2000          \\
$\alpha$ Cet & 2  & 2 &   11.1 $\pm$ 0.8 &   ~~~586 $\pm$ 47  \\
$\sigma$ Lib & 2  & 0 &   ~7.5 $\pm$ 0.3 &   $>$1500  \\
$\lambda$ Aqr & 2 & 0 &  ~7.9 $\pm$ 1.4 &   $>$1000   \\
$\beta$ Peg  &  3 & 0 &  ~7.7 $\pm$ 0.5 &   $>$2500   \\
$\tau^4$ Eri &  1 & 1 &    12.4 $\pm$ 0.3 &   ~~~687 $\pm$ 14   \\
$\mu$ Gem    & 6  & 1 &  10.5 $\pm$ 1.2  &   ~~~798 $\pm$ 73   \\
$\delta$ Vir & 4  & 0 &  12.3 $\pm$ 1.2 &  $>$2500   \\
10 Dra       & 3  & 3 &  14.8 $\pm$ 1.6 &  ~~~151 $\pm$ 11   \\
$\rho$ Per   & 2  & 0  &  ~9.7 $\pm$ 1.0 &  $>$1000  \\
BS6861       & 1  & 0  &  48.5 $\pm$ 2.9 &  $>$1000   \\
$\delta^2$ Lyr & 2 & 2 & 16.2 $\pm$ 1.5 & ~~~465 $\pm$ 41  \\
RR UMi       & 1  & 0  &10.0 $\pm$  0.8 &  $>$2000    \\
$\alpha$ Her & 2 & 4 & 11.1 $\pm$ 0.7 & ~~~102 $\pm$ ~8  \\
OP Her       & 2 & 1 & 11.3 $\pm$ 1.2 & ~~~329 $\pm$ 31  \\
XY Lyr       & 1 & 3 &  15.0 $\pm$ 0.4 & ~~~223 $\pm$ 16    \\
R Lyr        & 1 & 3 &~6.4 $\pm$ 0.3 & ~~~368 $\pm$ 44   \\
RZ Ari       & 3 & 2 &~7.9 $\pm$ 0.8 & ~~~607 $\pm$  48  \\
30g Her      & 2 & 3 & 12.5 $\pm$ 1.1 & ~~~211 $\pm$ 42   \\
SW Vir       & 2 & 2 & 22.0 $\pm$ 4.7 & ~~~432 $\pm$ 37  \\
RX Boo       & 0 & 3 &    ~~~...       & ~~~233 $\pm$  20  \\
\noalign{\smallskip}
\hline \hline
\noalign{\smallskip}
\end{tabular}
\vspace{-3mm}
\begin{list}{}{}
\item[$^{\mathrm{a}}$]   number of weak lines of $^{13}$C$^{16}$O. 
\item[$^{\mathrm{b}}$]   number of weak lines of $^{12}$C$^{17}$O. 
\end{list}
\vspace{-3mm}
\end{table}

\begin{figure*}
\hspace{5mm}
\includegraphics[width=16.0cm]{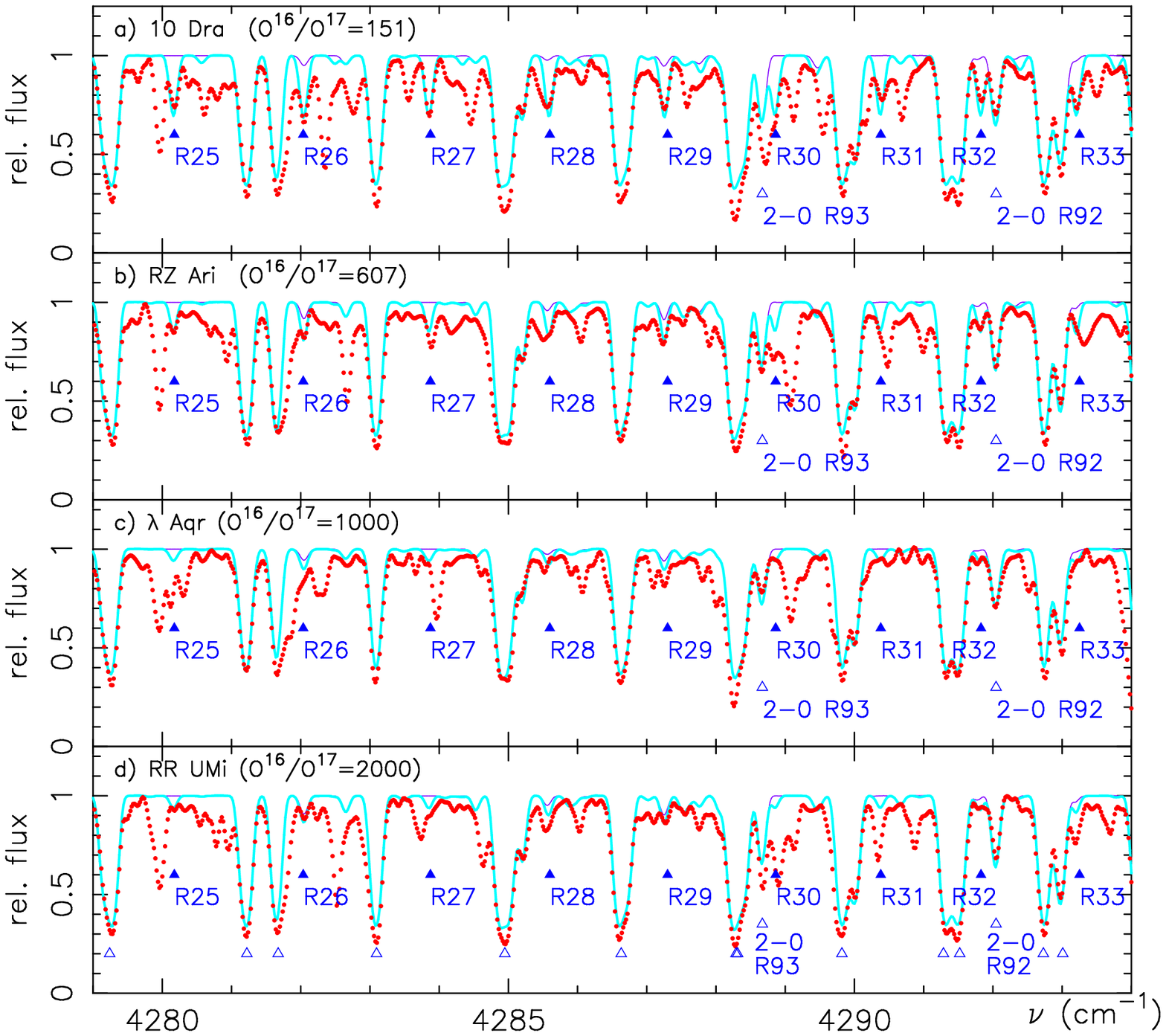}
\caption{ Observed spectra are shown  by  dots. Some $^{12}$C$^{16}$O 
 and $^{12}$C$^{17}$O features are indicated by open and filled
triangles, respectively. Synthetic spectra with assumed
 $^{16}$O/$^{17}$O ratios are shown by heavy solid lines while
those with null $^{17}$O by light solid lines.
{\bf a)} Observed spectrum of 10 Dra  compared with the synthetic one with 
$^{16}$O/$^{17}$O = 151. 
{\bf b)} Observed spectrum of RZ Ari compared with the 
synthetic one with $^{16}$O/$^{17}$O = 607. 
{\bf c)} Observed spectrum of $\lambda$ Aqr compared with the synthetic 
one with $^{16}$O/$^{17}$O = 1000.  
{\bf d)} Observed spectrum of RR UMi compared with the synthetic one with 
$^{16}$O/$^{17}$O = 2000.
}
\label{fig16.eps}
\end{figure*}

\subsection{Synthetic spectra}

We apply the spectral synthesis method as a check of
the results given in Sect.\,6.1. With the carbon abundance
and micro-turbulent velocity from Table 6 and with the isotopic
ratios from Table 7, 
we generate synthetic spectra between 4279 and 4294 cm$^{-1}$,
where $^{12}$C$^{17}$O (2-0) $R\,25 - 33$ lines are
found. Before we compare the synthetic spectra 
with observed ones,  we corrected for the effects of the
instrument broadening  with the Norton-Beer's apodization function 2 
(Norton \& Beer 1976), which was used in apodization of the observed 
spectra (see Paper I),
and also of the macro-turbulent broadening based on the measured 
intrinsic line widths (see Table 4 of Paper III)\footnote{
The macro-turbulent velocity is derived by subtracting the effect
of the micro-turbulent velocity from the line width on the 
assumption of Gaussian micro- and macro-turbulence. The resulting 
values are; $\xi_{\rm macro}$ = 2.7, 2.1, 3.9,
and 2.8 km\,s$^{-1}$ for 10 Dra, RZ Ari, $\lambda$ Aqr, and RR UMi,
respectively.}.  
As an example, we compare the observed spectrum of 10 Dra (dots) with 
the synthetic one (heavy solid line) in Fig.\,16a. In this region, the weak 
line of $^{12}$C$^{16}$O  (2-0) $R$\,92 can clearly be seen at 4292 cm$^{-1}$,
which serves as a check of the carbon abundance
\footnote{Strong lines noted by the open triangle in Fig.\,16d are mostly
originating from  $^{12}$C$^{16}$O (3,1) and they are too strong to be
used as indicator of the carbon abundance.}.   
The lines of $^{12}$C$^{17}$O (2-0) $R$\,25, 27, and 32 used to
determine $^{12}$C$^{17}$O abundance in Sect.\,6.1 can be 
reproduced well by the synthetic spectrum. 
Some other lines of $^{12}$C$^{17}$O such as (2-0) $R$\,26 and 29
should include extra components, since the synthetic 
spectrum computed without $^{12}$C$^{17}$O at all (light solid line)
shows some absorption features.
Also some $^{12}$C$^{17}$O lines are badly blended with other lines, but 
all the lines of $^{12}$C$^{17}$O are not contradicting with the synthetic 
spectrum computed for $^{16}$O/$^{17}$O ratio of 151 (Table 7).
As another example, the case of RZ Ari is shown in Fig.\,16b. Although
the  $^{12}$C$^{17}$O lines are weaker than in the case of 10 Dra, some
unblended features such as (2-0) $R$\,25 and 32 and all the other
lines of $^{12}$C$^{17}$O are not contradicting with the synthetic spectrum
computed for $^{16}$O/$^{17}$O ratio of 607 (Table 7).

The major purpose of computing the synthetic spectra, however,
is to estimate the $^{16}$O/$^{17}$O ratio for the cases in which the 
$^{12}$C$^{17}$O lines are too weak for {\it EW}s to be measured accurately.
For example, all $^{12}$C$^{17}$O features are too weak for
their {\it EW}s to be measured in the case of $\lambda$ Aqr shown in Fig.\,16c
and these features, including relatively undisturbed ones such as
(2-0) $R$\,32 and 33, do not contradict with the synthetic spectrum
computed for $^{16}$O/$^{17}$O ratio of 1000. As another example,
the observed spectrum of RR UMi shown in Fig.\,16d may not contradict 
with the synthetic spectrum computed for $^{16}$O/$^{17}$O ratio 
of 2000. This conclusion depends largely on subtle depressions in
(2-0) $R$\,25 and 32. On the other hand, definite absorption 
seen at the position of $R$\,26 and 29 also appear in the synthetic 
spectrum computed without $^{12}$C$^{17}$O at all (light solid line)
and are not due to $^{12}$C$^{17}$O. However, it is possible
that the features assumed to be due to $^{12}$C$^{17}$O may still
include unknown impurities, and the $^{16}$O/$^{17}$O ratio estimated
on such subtle features should at best be regarded as a lower estimate
to the $^{16}$O/$^{17}$O ratio. The $^{16}$O/$^{17}$O ratios estimated
this way are given in the 5-th column of Table 7 without error bars.

\subsection{Uncertainties and Comparisons with other authors}

 The serious uncertainties in the analysis of such faint lines as
$^{12}$C$^{17}$O in cool stars are the location of the continuum
level and effect of blending by many weak lines. As a guide to
examine such problems, we computed synthetic spectra such as 
Fig.16 for the regions with rich lines of $^{12}$C$^{17}$O 
(4264 - 4295\,cm$^{-1}$) as well as of $^{13}$C$^{16}$O (4235 - 
4266\,cm$^{-1}$)  for all the spectra  analyzed. 
This was of some help to reject blending lines in our 
line list. Further, we examined the correlations among line 
intensities, line-widths, line-depths, 
and radial velocities, and rejected those lines that show large
deviations. The continuum level was estimated by connecting the
highest peaks by a smooth curve and the consistency of the resulting
 continuum can also be checked with the synthetic spectra. However,
we must look for the spectra with still higher resolution to
have final check of the adopted continuum. Also, a comparison
of the different authors' results (see below) suggests that
the errors in the final isotopic ratios due to such uncertainties 
may be as large as 100\% (or factor of two).    

The carbon isotopic ratios were studied relatively well
and we compare our  $^{12}$C/$^{13}$C ratios
with those by Smith \& Lambert (1990) and by other authors in Table 8.
 In addition to those cited in
Table 8, Wallerstein \& Morell (1994) reported the $^{12}$C/$^{13}$C 
ratios for 8 early M giants to be 7 - 12, although no object
common to our sample is included. 
Our results on $^{12}$C/$^{13}$C ratios appear to agree rather well
with the results by other authors in general.

The determination of the oxygen isotopic ratios is more difficult 
but a detailed identification of $^{12}$C$^{17}$O features was 
done on the high resolution FTS spectra  by Maillard (1974), who estimated
$^{16}$O/$^{17}$O ratio in $\alpha$ Her for the first time.
We compare our  $^{16}$O/$^{17}$O ratios with  those by Harris \&
Lambert (1984) and by other authors in Table 8. Also, Wallerstein 
\& Morell (1994) reported
lower limits to $^{16}$O/$^{17}$O ratio in 8 early M giant stars to be
85 - 295. Inspection of Table 8 reveals that $^{16}$O/$^{17}$O ratios 
by different authors differ by factors of 2 - 4. This may largely 
be due to difficulty to locating the continuum levels and to measuring
weak features,  and there should be
a room to improve the results by better spectroscopic data.

\begin{table*} 
\centering
\caption{Comparisons with the results by other authors }
\vspace{-2mm}
\begin{flushleft}
\begin{tabular}{ r c c l l c c l l}
\hline \hline
\noalign{\smallskip}
 Obj.  & log $A_{\rm C}^{~~a}$ & log $A_{\rm C}$ & log $A_{\rm O}^{~~a}$
 & log $A_{\rm O}$ &  $^{12}$C/$^{13}$C$^{~a}$ &
$^{12}$C/$^{13}$C & $^{16}$O/$^{17}$O$^{~a}$ &   $^{16}$O/$^{17}$O\\
\noalign{\smallskip}
\hline
\noalign{\smallskip}
$\alpha$ Tau & 
8.38 $\pm$ 0.04 & 8.40 $\pm$ 0.07$^b$ & 8.78 $\pm$ 0.07 & 8.78 $\pm$ 0.08$^b$ &
11 $\pm$ 1  &  10 $\pm$ 2$^b$ & $>$1000  & ~560 $\pm$ 180$^f$ \\
    &            &  8.35 $\pm$ 0.20$^c$ &               & 8.93 $\pm$ 0.20$^c$ &
            &   10 $\pm$ 2$^c$         &               &             \\    
$\nu$  Vir   & 
8.13 $\pm$ 0.08 & 8.51 $\pm$ 0.08$^b$ & 8.84 $\pm$ 0.05 & 8.91 $\pm$ 0.06$^b$ &
~9 $\pm$ 1  &   12 $\pm$ 2$^b$  &      & \\
$\alpha$ Cet  & 
8.64 $\pm$ 0.05 & 8.20 $\pm$ 0.30$^c$ & 8.98 $\pm$ 0.14 & 8.93 $\pm$ 0.30$^c$& 
  11 $\pm$ 1  &  10 $\pm$ 2$^c$ &   &   \\
$\beta$ Peg  & 
8.27 $\pm$ 0.05 & 8.42 $\pm$ 0.07$^b$ & 8.77 $\pm$ 0.12 & 8.81 $\pm$ 0.06$^b$ &
  ~8 $\pm$ 1  &  ~8 $\pm$ 2$^b$ &       $>$2500 & 1050 $\pm$ 375$^f$\\
            &
            &  8.20 $\pm$ 0.40$^c$ &                  & 8.93 $\pm$ 0.40$^c$ &
              &   ~5 $\pm$ 3$^c$    &                    &                \\
$\mu$ Gem & 
8.32 $\pm$ 0.06 & 8.43 $\pm$ 0.04$^b$ & 8.81 $\pm$ 0.13 & 8.82 $\pm$ 0.12$^b$ &
11 $\pm$ 1 &  13 $\pm$ 2$^b$ & ~~~798 $\pm$ 73 &  ~325 $\pm$ 112$^f$ \\
$\delta$ Vir & 
8.50 $\pm$ 0.06 & 8.61 $\pm$ 0.10$^b$ & 8.84 $\pm$ 0.15 & 8.82 $\pm$ 0.06$^b$ &
12 $\pm$ 1  &  16 $\pm$ 4$^b$ &      &  \\
10 Dra  & 
8.43 $\pm$ 0.14 & 8.59 $\pm$ 0.08$^b$ & 8.92 $\pm$ 0.18 & 9.04 $\pm$ 0.09$^b$ &
15 $\pm$ 2 &  12 $\pm$ 3$^b$ &   &  \\
$\rho$ Per & 
8.27 $\pm$ 0.07 & 8.46 $\pm$ 0.04$^b$ & 8.87 $\pm$ 0.07 & 8.92 $\pm$ 0.15$^b$ &
10 $\pm$ 1  &  15 $\pm$  2$^b$ &   &  \\
$\alpha$ Her &   
   &        &     &        & 11 $\pm$ 1  & 17 $\pm$ 4$^d$    & 
~~~102 $\pm$ ~8 & ~190 $\pm$ 40$^f$ \\
        &
   &       &     &        &        & ~5 $\pm$ 1$^e$    & 
 & ~450 $\pm$ ~50$^e$ \\         
30g Her & 
8.35 $\pm$ 0.11 & 8.25 $\pm$ 0.07$^b$ & 8.75 $\pm$ 0.09 & 8.73 $\pm$ 0.08$^b$ &
13 $\pm$ 1 &   10 $\pm$ 2$^b$ & ~~~211 $\pm$ 42 & ~675 $\pm$ 175$^g$  \\
\noalign{\smallskip}
\hline \hline
\noalign{\smallskip}
\end{tabular}
\end{flushleft}
\vspace{-6mm}
\begin{list}{}{}
\item[$^{\mathrm{a}}$]   present results.
\item[$^{\mathrm{b}}$]   Smith \& Lambert (1990). 
\item[$^{\mathrm{c}}$]   Decin et al. (2003).
\item[$^{\mathrm{d}}$]   Hinkle  et al. (1976).  
\item[$^{\mathrm{e}}$]   Maillard (1974).  
\item[$^{\mathrm{f}}$]   Harris \& Lambert (1984).
\item[$^{\mathrm{g}}$]   Harris et al. (1985).  
\end{list}
\vspace{-3mm}
\end{table*}

\section{Discussion}

\subsection{Abundance analysis and photospheric structure }

The abundance analysis of very cool stars
is associated with many difficulties. In particular, absorption lines tend 
to be strong due to  the low continuous opacity at low temperatures
of  the photospheres of cool giant stars. The consequences are severe blending
and uncertainties in the continuum level, which are not only the obvious
difficulties by themselves but  also make it difficult to measure
other weak lines which are most important in abundance determination.
In fact, if a sufficient number of weak lines which are linearly
proportional to the effective column density of a species under
consideration can be measured, abundance determination is quite
easy. Such an ideal case is seldom realized especially in cool
stars. However, it was thought  possible to carry out an
abundance analysis with the use of the micro-turbulent model 
(or curve-of-growth method) even if saturated lines had to be used. 
This model has in fact been applied with reasonable success to  
abundance analysis in general until recently. 

The real difficulty in the abundance analysis of cool stars is that the
classical method based on the micro-turbulent model can
no longer be applied for lines of equivalent width larger than a 
certain critical value, which we found to be log $W/\nu \approx -4.75$
for the case of M giant stars (Sects.\,4 \& 5).  Since stronger lines are
formed in the upper photospheric layers where modeling may be
increasingly uncertain, this difficulty can be related to the
problem of the  model photospheres. Previously,
Smith \& Lambert (1990)  noted that the CO lines stronger
than about log $W/\nu \approx -4.8  \sim -4.7$ should not be
used for abundance analysis, since these lines are formed in the
layers above the continuous optical depth of $ \tau_{c} \approx 10^{-3} $,  
where the photospheric structure cannot be described well by the usual
model photospheres. We arrive at a similar conclusion that the
lines stronger than log\,$W/\nu \approx -4.75$  cannot be used for 
abundance analysis, but by somewhat different argument based on 
purely empirical evidence shown most clearly for the OH lines 
of the fundamental bands (Fig.\,10).

Then, a problem is again why the intermediate-strength lines show such
abundance corrections as large as 1.0\,dex or more when the weak lines 
show null corrections? Probably, it is unlikely 
that the uncertainty in the thermal structure of the upper layers    
could produce such a large effect on the intermediate-strength lines, 
since differences in the model photospheres have only modest effects on 
abundances as we have seen before (Sects.\,4.3 \& 5.3). Also, no model
of the upper photosphere  to explain the intermediate-strength
lines or the stronger lines is known. On the other hand, dynamical 
structure represented by stellar turbulence may have more serious 
effect on the intermediate-strength lines  heavily saturated.
 Taking advantage that the line  positions and line profiles
accurately can be measured with FTS, we showed that the line asymmetries 
and differential line shifts of CO lines are quite large and that a  
simple model of the depth-independent and isotropic Gaussian 
micro-turbulence cannot be justified (Paper III). 

Even if we know that a simple model of micro-turbulence is not 
sufficiently realistic enough, yet it is unknown how the behavior of the 
intermediate-strength lines (Figs.\,4 - 11 \& 13 - 14) can be reproduced with
better treatment of stellar turbulence. In this connection, it is 
interesting if a more consistent treatment of the dynamical 
structure of stellar photosphere by recent 3 D models 
can resolve the present difficulty. Recently, a detailed
numerical simulation of surface convection in red giant stars is
reported by Collet et al.\,(2007). Although their simulations
were not extended to the regime of M giants, their result
for the coolest model of $T_\mathrm{eff} = 4697$\,K and 
[Fe/H] = 0.0 indicated that  a difference between 3D and 1D  oxygen abundances
derived from OH lines (not ro-vibrational lines but of A - X system
at 3150\,\AA) is about 0.1 dex.  This seems to be
due to the difference of the thermal structures between the 3D and 1D
models, but possible effect of the dynamical structure of the
3D models on the saturated lines is unknown yet.  
Their results showed that the effect of the 3D structure should
be more important in stars of the lower metallicities, but 
it is unclear if such a large difference in abundances derived 
from the weak and intermediate-strength lines in our population I 
sample can be relaxed by considering the dynamical effect of convection 
in the 3 D models. Thus we are inclined to conclude that the difficulty
we have encountered may not be resolved within the framework
of the photospheric structure.  

\subsection{Abundance analysis and extra molecular envelope}

We already know that the strong lines of CO in late M giant stars
are badly disturbed by the contributions of the extra molecular
layers (Paper II).
Now, since the intermediate-strength lines and the strong lines
behave similarly on the
$ \Delta$\,log\,$A$ - log\,$W/\nu$ diagram (e.g. Figs.\,4 - 5), an 
interesting question arises if the intermediate-strength lines  can also be
understood similarly as the strong lines. For the late M giants, 
the intermediate-strength lines
are not necessarily originating from the levels of low LEP (see Fig.\,12a),
but still increasingly reduced contributions from higher excitation
lines can be possible for modest kinetic temperatures of the warm
molecular layers, and the trend of the intermediate-strength lines
together with the strong  lines in these giants (Figs.\,4 - 5) can be 
explained consistently. In fact, given that the presence of the 
MOLsphere has been demonstrated in the late M giants (Paper II), 
it should at the same time produce excess
absorption in the intermediate-strength lines as it
does for the strong lines.  It is then quite natural to assume
the same origin for the intermediate-strength lines and the strong
lines in the late M giants.     

Interestingly, the intermediate-strength lines of CO in the early M 
giants (including a K giant) mostly give larger abundance corrections 
compared with those from the weak lines (e.g. $\beta$ Peg, $\alpha$ Tau, 
$\nu$ Vir, $\alpha$ Cet in Figs.\,6 - 7).
So far, we have not considered  the effects of outer molecular 
layers in these  early M giant stars\footnote{We noticed that the 
lines of log\,$W/\nu > -4.5$ show excess absorption in some early M 
giants in Paper II, but we regret that we did not pursue this case 
in more detail. With the critical value of log\,$W/\nu \approx -4.5$, 
the lines showing excess absorption were limited to only few lines 
in the early M giant stars.}. 
But why not?  Especially, the behaviors of the intermediate-strength
lines on the $ \Delta$\,log\,$A$ - log\,$W/\nu$ diagrams are so similar 
for the late M (Figs.\,4 - 5) and  early M (Figs.\,6 - 7) giants, 
it is natural to assume the same origin for the intermediate-strength 
lines in the early and late M giants. Furthermore, we  
identified  H$_2$O bands on the ISO spectra  of the K - early M
giant stars included in our present sample (e.g. $\alpha$ 
Tau, $\beta$ Peg, and $\alpha$ Cet) and one possibility is that these  
H$_2$O bands are originating in the outer molecular layers as for the 
late M giants (Tsuji 2001). But this is not a unique interpretation:
For example, Ryde et al.\,(2002) detected H$_2$O lines on the high resolution 
spectrum of the K giant $\alpha$ Boo and suggested that its origin can be
due to anomalous structure of the photosphere. We would like to point
out, however,  that the anomalous behavior of the intermediate-strength
lines and the presence of H$_2$O bands in the K - early M giant stars
can consistently 
be understood by assuming the presence of the outer molecular layers.

In the case of CO lines in late M giants such as $\rho$ Per, R Lyr, 
and SW Vir, and also in the M3 giant  $\tau^4$ Eri,  the 
intermediate-strength lines and the strong lines show quite 
different behaviors as shown in Figs.\,8 - 9:
The strong lines  show positive abundance corrections 
while the intermediate-strength lines show mostly negative  corrections
for the carbon abundances obtained from the weak lines.
We first thought that this difference could be a reason why we should
consider different explanations for the intermediate-strength lines
and the strong lines. However, we soon arrive at an opposite 
conclusion that this fact can be
supporting evidence for the same origin  of the intermediate-strength
lines and the strong lines. In fact, a possible contribution from the
outer molecular layers can be either absorption or emission
depending on the extension, temperature, and optical thickness
of the outer molecular layers for a particular line. We can 
therefore explain  the negative abundance corrections for the 
intermediate-strength lines  by possible reductions of their {\it EW}s
due to filling in by the emission from the  outer molecular layers.
On the other hand,  we have presently no explanation for the 
intermediate-strength lines if we assume photospheric origin for 
them. 

Finally, a problem is whether the intermediate-strength lines of OH  can be 
understood in the same way as for the CO lines. Unlike CO, OH may not be so
abundant in the outer molecular layers, since oxygen may be  mostly in 
H$_2$O at relatively cooler environment expected 
for the MOLsphere. For this reason, we first thought that the anomalous 
strengths of the intermediate-strength lines of OH  may not be due to 
contamination from the outer molecular envelope, but may instead be 
originating in the photosphere itself. 
However, the infrared OH lines shows behaviors quite similar
to the CO lines (e.g. Fig.\,4 \& 10), and it is again most natural to assume
the same interpretation for the OH and CO lines. Given that the presence
of OH is known even in the cooler circumstellar envelopes of red giant stars
by OH masers, it may be  possible that OH can be abundant in the
warmer MOLsphere. Furthermore the assumption of the chemical equilibrium 
should probably not be applied to the outer layers and we should
rather use the OH lines as probes of non-equilibrium processes in the
outer layers. Then, we conclude that the intermediate-strength lines of
OH  observed in the late M giants such as $\alpha$ Her, R Lyr, and SW vir 
(Fig.\,10 - 11) as well as in the earlier M giant such as $\sigma$ Lib
(Fig.\,14b) also include  contributions from the outer molecular layers. 

\begin{figure*}
\centering
\includegraphics[width=16.0cm]{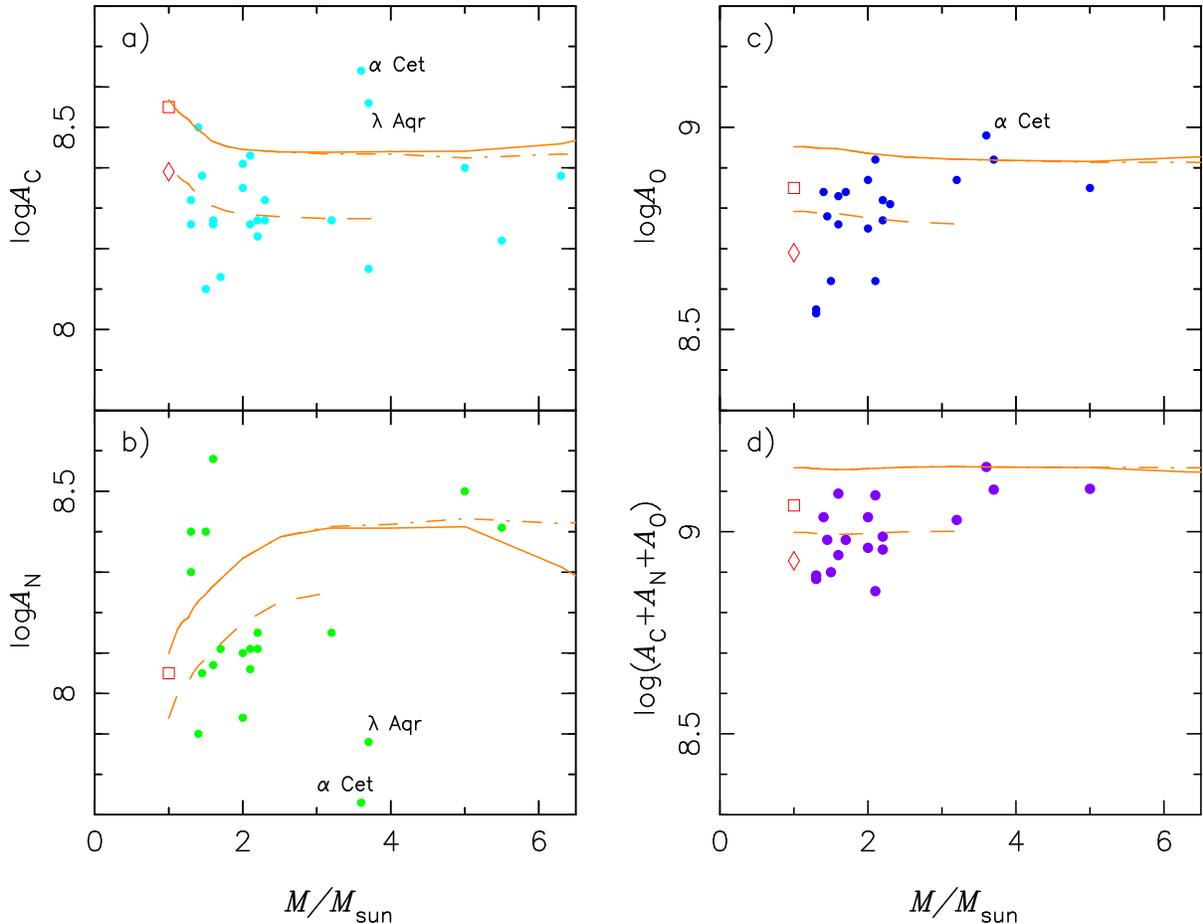}
\caption
{Observed abundances of {\bf a)} carbon (Table 6), {\bf b)} nitrogen
(Aoki \& Tsuji 1997; Table 6), {\bf c)} oxygen (Table 6; If the results 
derived from the $H$ and $L$ band spectra are available, mean values 
are used), and {\bf d)} C+N+O,  compared with the predictions of 
recent evolutionary models (Claret 2004). The results of FDU and SDU for 
the case of $Z = 0.02$ 
are shown by solid and dash-dotted lines, respectively, in each panel. 
The observed C + N + O abundances are smaller by about 0.16 dex
compared with the predictions based on $Z = 0.02$ as can be seen
in panel d. This initial metallicity correction of -0.16 dex is applied 
to the predicted C, N, and O abundances in panels a, b, and c, respectively,
and the corrected results are shown by dashed lines. In each panel, 
the solar abundances are shown by open square (Anders \& Grevesse 1989; 
Ayres et al. 2006) and by open diamond (Allende Prieto et al. 2002).
}
\label{fig17.eps}
\end{figure*}

The possible presence of the outer molecular layers in red giant stars
is supported by direct measurements of the angular extensions. 
For example, angular diameters of normal K and M giant stars, including
several objects common to our sample, were measured
in the strong TiO bands at 712\,nm and in the nearby 
pseudo-continuum at 754\,nm (Quirrenbach et al. 1993).  The observed
712\,nm/754\,nm diameter ratios appeared to be larger than
predicted by model photospheres,  suggesting that the
photospheric models cannot describe adequately the outer layers
of these red giant stars,  and the result may lend support for the presence of 
the extra molecular envelope (Quirrenbach 2001). 
Also, recent interferometric observations in the infrared bands provided 
direct support for the presence of extended warm molecular gas layers not
only in Mira variables but also in the late M giants (e.g.
Mennesson et al. 2002; Perrin et al. 2004). 
On the other hand, the apparent angular diameter of $\alpha$ Her
in the mid infrared was found to be larger by 30\% than its
near infrared size, but the high resolution 11\,$\mu$m spectrum does not
show any substantial spectral lines in $\alpha$ Her (Weiner et al. 2003).
This result was explained by a balance
of absorption and  emission in a spherically extended envelope, and
the mid infrared spectra and visibilities could be consistent with 
the warm molecular envelope model (Ohnaka 2004).

It is to be remembered that a possible  presence of the warm molecular 
envelope in the early M giant stars was previously noticed when we first 
identified H$_2$O bands on the spectrum of the M2 giant $\beta$ Peg
observed  with ISO SWS (Tsuji et al. 1997). Also, H$_2$O bands were
found in several M giants earlier than M6 by Matsuura et al.(1999) with IRTS
(Infrared Telescope in Space) launched in March 1995 by ISAS.   
These results were further extended to several K - M giant stars (Tsuji 
2001) and also confirmed by other authors (e.g. Decin et al. 2003). The
implication of this finding, however, was not necessarily
realized very well even by ourselves, and it took some time before 
we recognize that this finding should also be related to  
the difficulty encountered in the abundance analysis of red giant stars. 
Since the temperatures and radial velocities of the MOLsphere differ
only slightly  from those of the upper photosphere, it was 
difficult to recognize the presence of the MOLsphere by the
spectroscopic method\footnote{ From a detailed measurement of the 
shifts and shapes of CO lines, we showed that the line asymmetries and 
differential line shifts of CO lines are as large as 0.5\,km\,s$^{-1}$
(Paper III). This results can also be interpreted as 
evidence for the presence of the extra molecular layers that may have 
some velocity differences against the photosphere.}. 
However, we finally recognize the effect of the overlaying warm 
molecular layers  on the high resolution infrared spectra of dozens of 
red giant stars. 

Considering several lines of reasonings outlined above, we conclude 
at last that the presence of a rather warm and quasi static molecular 
envelope above the photosphere is a common feature in K and M giant stars. 
It is quite unexpected and even surprising if both the CO and OH 
lines of intermediate-strength with LEP as high as 2.0 eV (Fig.\,12) include
the absorption and/or emission originating in the outer molecular layers.
We have previously noticed such a possibility mainly for the late M
giants (Paper II), but now we find that the presence of the extra molecular
envelope is a general phenomenon not only 
in the late M giants but also in the early M and K giants as well. 
 Also, the fact that the rather high excitation lines are included in
these intermediate-strength lines implies that the extra molecular 
envelope is indeed ``warm'' and should be even warmer than we
imagined before from the analysis of the strong lines (Paper II).
Yet we know little about this new feature consisting of warm gaseous
molecules such as CO, OH, and H$_{2}$O.
We hope it  our next step to extend our analysis as done for the
strong lines (Paper II) to the intermediate-strength lines and to 
clarify the nature of the warm molecular layers in more detail.

We conclude that the observed infrared spectra should actually be a
hybrid of multiple components originating in the photosphere and
overlaying warm molecular layers or MOLsphere. The hybrid nature of the 
infrared spectra of red giant stars is now  confirmed both in the 
low resolution ISO spectra and high resolution FTS spectra. 
We must then abandon the traditional idea that the stellar spectra 
are primary defined by  the stellar photospheric structure, at least  
for the infrared spectra of K - M giant stars.

\subsection{CNO abundances }

Even if the photosphere of a cool giant star is veiled by
the warm molecular envelope, the weak lines of high excitation may
remain almost undisturbed\footnote{We also assume that the MOLsphere
is almost transparent to the continuum radiation at least in the near
infrared and that the infrared flux method with the $L$ band flux 
can still be applied. However, it is possible that the MOLsphere
will be opaque to the continuum radiation at the longer wavelength region
as has actually been shown for the radio domain of Mira variable stars
(Reid \& Menten 1997). }. In fact the weak lines  are clearly
distinguished  from the intermediate-strength lines on the
$\Delta$\,log\,$A$ - log\,$W/\nu$ diagram (Figs.\,4 - 11, 13 - 14),
and we confirmed that the weak lines  well behave as
expected from the classical micro-turbulent model (Sects.\,4 \& 5).
Thus, we are convinced that  abundance determination can be possible 
with the use of the weak lines of log $W/\nu \la -4.75$.  With this caution,
our results agree well with those by Smith \& Lambert (1990), who  
already used only weak lines in their abundance analyses. Then we
compare our results with the predictions of recent evolutionary models.  

We plot the resulting C and O abundances in Figs.\,17a and 17c,
respectively. Also, we plot N abundances based on our
previous analysis of the same sample (Aoki \& Tsuji 1997),
corrected for the effect of the revised C abundances (14-th column
of Table 6), in Fig.\,17b. For 5 late M giants for which the N 
abundances based on the ro-vibrational lines of NH are available,
we plotted the mean log\,$A_{\rm N}$ values based on the lines 
of CN and NH.  Given C, N, and O abundances, we obtain
the sum of the C, N, and O abundances 
as plotted in Fig.\,17d. The results are compared with the predictions
of the standard evolutionary models by Claret (2004): The
results of the first dredge-up (FDU) are shown by solid lines 
and those of the second dredge-up (SDU)  by  dash-dotted lines
in Figs.\,17a-d.  

We first examine the sum of the C, N, and O abundances in Fig.\,17d. 
The model prediction simply confirms that the CNO processing
conserved the sum of the C, N, and O abundances corresponding to
$Z = 0.02$, as assumed in the evolutionary models. The observed
values are generally smaller than the values of the models which are 
log ($ A_{\rm C} + A_{\rm N} + A_{\rm O}) \approx 9.16 $. The mean 
of the observed values in low mass stars ($ M < 2.5 M_{\sun} $)  is 
log ($ A_{\rm C} + A_{\rm N} + A_{\rm O}) \approx 9.00 $ as shown by 
the dashed line in Fig.\,17d, and thus
the mean metalliciy of the low mass stars we analyzed
is smaller than the value assumed in the evolutionary models by
Claret (2004). This  metallicity difference should be
kept in mind in comparing the observed and predicted
C, N, and O abundances. Also, a recent solar value of the C+N+O
abundances based on the C and O abundances by Alende Prieto et al. 
(2002) is shown by 
the open diamond and a classical value (e.g. Anders Grevesse 1989), 
recently supported by Ayres et al. (2006), is indicated by the open squares 
in Fig.\,17d. The N abundance is taken from  Anders \& Grevesse (1989).

The observed carbon abundances are generally smaller than those of the
model predictions, but may be consistent with the prediction after the
metallicity difference between the evolutionary models and actual stars
is corrected as shown by dashed line in Fig.\,17a. The scatters
around the dashed line may be due to those of the  metallicity in
our sample, except for a few objects. For example, two stars appear
above the predicted FDU curve: The object showing the largest deviation 
is $ \alpha $ Cet  and the other one is $ \lambda $ Aqr. It is 
interesting to notice that these two stars also show the largest under 
abundances of N at $ M \approx 3.6 M_{\sun} $ in Fig.\,17b .

The nitrogen abundances may roughly be consistent with the FDU curve 
with and/or without metallicity correction in general. However, 
the deviations from the prediction appear to be larger in a few
objects, and it is as if some extra mixing may be needed as in the
case of the $^{12}$C/$^{13}$C ratios discussed in Sect.\,7.4.
Most lines of both CN and NH used are weak lines that satisfy
the condition of log\,$ W/\nu \la -4.75 $, and N abundances may be free
from the difficulty encountered in the analysis of the CO first overtone
bands in Paper I (also Sect.\,4.1).  
 
Finally, the oxygen abundances in Fig.\,17c show a similar pattern as 
the C+N+O abundances in Fig.\,17d, and the scatters may represent those of the 
metallicity in our sample. There is a tendency for the oxygen abundances
to be larger in higher mass objects. The resulting oxygen and carbon abundances
correlate rather well and the mean $A_{\rm C}/A_{\rm O}$ ratio for 19
objects for which C and O abundances are determined is 0.37 $\pm$ 0.10
(std. dev.).   

\begin{figure}
\centering
\includegraphics[width=8.0cm]{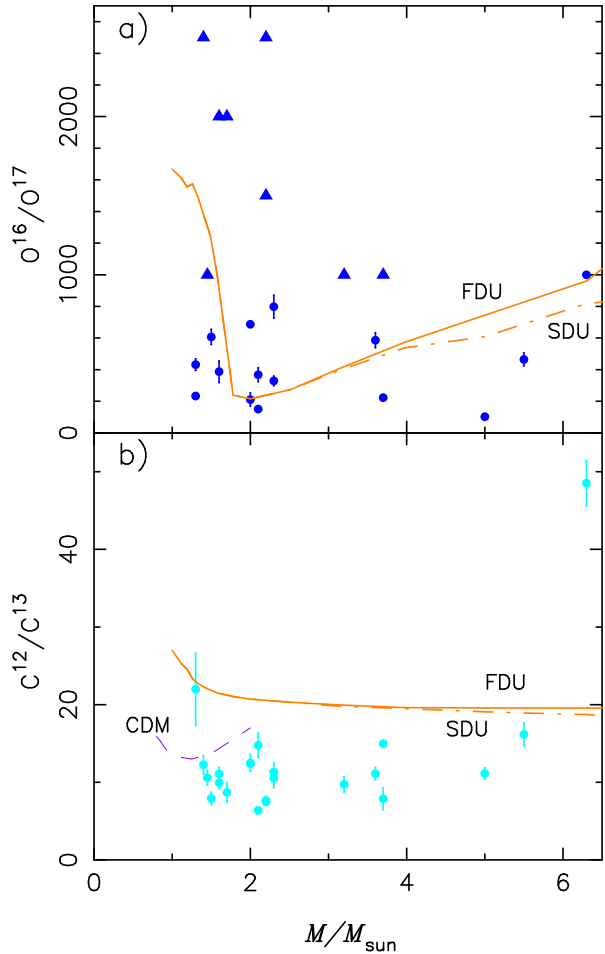}
\caption{ 
{\bf a)}  The observed $^{16}$O/$^{17}$O ratios are plotted 
against stellar mass by filled circles and by the filled triangles
if only lower limits can be estimated. Predicted $^{16}$O/$^{17}$O
ratios by FDU and SDU are shown by solid and dash-dotted lines,
respectively.
{\bf b)} The observed $^{12}$C/$^{13}$C ratios are plotted by
filled circles against stellar mass. Dashed curve is the
predicted $^{12}$C/$^{13}$C ratios by the ``Compulsory Deep Mixing
(CDM: Eggleton et al. 2008)''. 
}
\label{fig18.eps}
\end{figure}

\subsection{Carbon and oxygen isotopic ratios}

The resulting  $^{16}$O/$^{17}$O ratios (Table7) are 
plotted against the stellar mass in Fig.\,18a by filled circles
or by filled triangles if only lower limits could be estimated.
For comparison, predicted results of FDU and SDU  by Claret (2004) 
are shown  by solid  and dash-dotted lines, respectively.
The rather large variation in $^{16}$O/$^{17}$O of low mass stars
is well consistent with the prediction of the evolutionary models, 
confirming  the previous analysis by Dearborn (1992) using the 
observed data known at that time. 
In the higher mass stars, some objects have quite low 
 $^{16}$O/$^{17}$O ratios, but the general trend that the
$^{16}$O/$^{17}$O ratios are confined to  lower values
can be  consistent with the predicted ones. Considering the 
possible uncertainties in the stellar masses, the observed and
predicted  $^{16}$O/$^{17}$O ratios can be regarded as not showing 
a serious contradiction.

The observed $^{12}$C/$^{13}$C ratios (Table 7) are shown
by filled circles and compared with the predicted ones (Claret 2004) 
in Fig.\,18b.  The observed $^{12}$C/$^{13}$C ratios are now around 10
in contrast to our previous $^{12}$C/$^{13}$C ratios which were mostly 
below 10 (Tsuji 2007), but yet too small compared with $^{12}$C/$^{13}$C\,
$\approx 20$ predicted by the standard evolutionary models.  This 
contradiction has been known for a long time and we simply confirm this 
long-standing puzzle. To resolve this dilemma, 
extra mixing of the CN-cycle processed material had to be assumed.
For example, a  deep circulation  below the bottom of the
convective zone, referred to as ``cool bottom processing'', 
was proposed ( Boothroyd \& Sackmann 1999).
More recently, it was shown by Eggleton et al. (2008) that a deep 
mixing mechanism should operate in all the low mass giants due to 
a molecular weight inversion resulting from $^{3}$He\,($^{3}$He, 
2p)\,$^{4}$He reaction in the layers just above the H-burning shell.     
The predicted $^{12}$C/$^{13}$C ratios by this ``Compulsory Deep Mixing 
(CDM)''  are shown by the dashed line in Fig.\,18b. Now, the new
prediction appears close to the upper boundary of the observed results, and 
discrepancies between observations and predictions are relaxed
if not perfectly. 

\subsection{The warm molecular envelope and mass loss from red giants}

Given that the presence of the warm molecular envelope or 
MOLsphere is a general phenomenon
in normal red giant stars, the next problem is to clarify the
nature of the MOLsphere in detail, and to understand the structure 
of the entire atmosphere extending beyond the photosphere.
An initial step towards this goal is opened by a detailed MHD 
simulation of the red giant wind by Suzuki (2007), who showed that
nearly static regions of $T \approx 10^3 - 10^4$\,K are formed 
at several stellar radii above the photosphere.
It is interesting that the mass loss from red giant stars with the
wind velocities far smaller than the surface escape velocity can
be understood consistently with his simulation in which the stellar
winds are effectively  accelerated from the quasi-static region  
at several stellar radii above the photosphere. 

The quasi-static region predicted by Suzuki's simulation can be 
identified with the warm molecular envelope that we confirmed  
in several observational studies discussed in Sect.\,7.2;  
we hope that the simulation of Suzuki can be extended to include spectral 
modeling so that more detailed confrontation between theory and
observations can be achieved. We include in Table 3 the 
measured data on the stronger lines together with those on the weaker 
lines used in the abundance analysis, and we hope that these data can be 
interpreted correctly with more sophisticated  models in the future.     
With such a detailed confrontation between theory and observations, 
we expect to have a consistent solution to the long-standing problem 
on the origin of mass-loss from red giant stars for the first time.

\section{Concluding remarks}

We initiated this study with a hope to clarify the nature of the difficulty 
encountered in our previous analysis of CO (Papers I \& III ), since 
we felt that something is still not well understood in abundance analysis of
cool giant stars. At first, we had to recognize that the
line spectra of cool giant stars are in fact highly complicated
and may consist of at least three different groups of lines that may
have different origins: The weak lines of log $W/\nu \la -4.75$, the
intermediate-strength lines of $ -4.75 < {\rm log} W/\nu \la -4.4$, and
the strong lines of log $W/\nu > -4.4$. Of these three groups, only the 
weak lines could be understood well on the basis of the classical 
micro-turbulent model of line formation.
On the other hand, it has been known that the  strong lines  
could not be understood as being of the photospheric origin alone but
should include an extra component originating in the outer molecular
layers (Paper II).

The most difficult case was the intermediate-strength lines. However,
we are finally led to conclude that the intermediate-strength
lines  are essentially the same as the strong lines. This means that
not only the low excitation strong lines but also medium strong lines
with LEP as high as 2\,eV (Fig.\,12)  are disturbed by the contamination 
originating in the outer molecular envelope,
which we refer to as MOLsphere for simplicity. This in turn
means that most of the dominant molecular lines observed in the
infrared spectra of K - M giant stars should be badly disturbed by 
the contamination from the extra component beyond the photosphere. 
At the same time, this finding provides further support for the
presence of the MOLsphere, on the basis of the high resolution spectra
of dozens of red giant stars. 
Thus, the difficulty of the abundance analysis we have encountered
is nothing but a warning that the spectra of cool luminous stars
cannot be analyzed by the conventional methods using the classical model
photospheres. Instead, the influence of the 
extra constituent of the atmosphere, which we tentatively referred to as
MOLsphere, should be quite essential not only in the strong 
low excitation lines but also in many weaker lines that dominate 
the infrared spectra.

Now we had to know that
the largest fault in our previous analysis (Paper I) was to have assumed
that the critical value discriminating the weak and the intermediate-strength 
lines is log $W/\nu \approx -4.5$ rather than -4.75. For this reason,
many lines which are disturbed by the contamination from the warm 
molecular envelope were included in the abundance analysis of Paper I. 
It should certainly be difficult to attempt abundance analysis on such
lines including the contamination of the non-photospheric origin 
by the use of a conventional model photosphere, and  the reason why abundance 
analysis of cool giant stars often encountered difficulties appeared to
be quite simple, at least partly. This result also implies
that it should be difficult in principle to apply low resolution
infrared spectra to abundance analyses of cool luminous stars, and
the hybrid nature of the infrared spectra reinforces the reason
why high resolution spectroscopy should be indispensable in abundance
analyses.

On the other hand, it is fortunate that the weak lines are relatively 
free from disturbance by extra component of non-photospheric origin
and we are convinced that the abundance analysis of cool giant
stars can be possible with these lines.
With a limited number of the weak lines, we could determine
the C, O, and their isotopic abundances with reasonable accuracy.
The resulting CNO and their isotopic abundances are roughly 
consistent with the predictions of the first and second dredges-up
during the red giant phase. We noticed, however, that there are some 
deviations from the evolutionary models especially in \element[][12]{C}, 
\element[][13]{C}, and \element[][14]{N}   
abundances in some stars, and it seems that some additional episodic events
may be needed to explain them. Since the CNO isotopes in the 
universe, especially \element[][13]{C}, \element[][14]{N}, and 
\element[][17]{O},  are largely supplied by red giant stars, 
further refinements of both abundance analyses and evolutionary  
modelings should be pursued.

In conclusion, it turns out that high resolution infrared spectra
include wealth of information not only on  photospheric 
abundances but also on the structure of the entire atmosphere extending 
beyond photosphere. For this reason, the infrared spectra must be
of hybrid nature with contributions from multiple 
components. In the interpretation and analysis of the infrared spectra of  
cool luminous stars, it is essential to keep their hybrid nature in mind.  

\begin{acknowledgements}
    I am most grateful to  Drs. K. H. Hinkle and S. T. Ridgway  for
generous support in observing the FTS spectra at KPNO, in applying the
KPNO archival data, and in analyzing the FTS spectra. I also thank
an anonymous referee for careful reading of the text and for helpful
comments. Data analyses were in part carried out on common use data 
analysis computer system at the Astronomy Data Center (ADC) of
NAOJ. This work was supported by Grant-in-Aid for Scientific Research
(C) no.17540213.
\end{acknowledgements}


\begin{thebibliography}{}

\bibitem[1997]{Abia97}
        Abia, C., \& Isern, J. 1997, \mnras, 289, L11

\bibitem[2008]{Abia08}
        Abia, C., de Laverny, P., \& Wahlin, R. 2008, 
        A\&A, 481, 161

\bibitem[2002]{Allende02}
        Allende Prieto, C., Lambert, D. L., \& Asplund, M. 2002,
        ApJ, 573, L137

\bibitem[1989]{Anders89}  
        Anders, E., \& Grevesse, N. 1989, Geochimica et Cosmochimica
        Acta, 53, 197

\bibitem[2000]{Aoki97}  
        Aoki, W., \& Tsuji, T. 1997, A\&A, 328, 186

\bibitem[2002]{Asplund05}
        Asplund, M. 2005, \araa, 43, 481

\bibitem[2006]{Ayres06}
        Ayres, T. R., Plymate, C., \& Keller, C. U. 2006,
        ApJS, 165, 618 

\bibitem[1988]{Bauschlicher88}
        Bauschlicher, C. W., Langhoff, S. R., \& Taylor, P. R. 1988, 
        ApJ, 332, 531

\bibitem[1980]{Bersanelli91}
        Bersanelli, M., Bouchet, P., \& Falomo, R. 1991, A\&A, 
	252, 854

\bibitem[1980]{Blackwell80}
        Blackwell. D. E., Petford, A. D., \& Shallis, M. J. 1980, A\&A, 
	82, 249 

\bibitem[1999]{Boothroyd99}
        Boothroyd, A., \& Sackmann, I. -J. 1999,  ApJ, 510, 232

\bibitem[1978]{Cerny78}
        Cerny, D., Bacis, R., Guelachvilli, G., \& Roux, F. 1978, J. Mol.
        Spectros., 73, 154

\bibitem[1983]{Chackerian83}
        Chackerian, C. Jr. \& Tipping, R. H. 1983, J. Mol. Spectrosc., 99, 431

\bibitem[2004]{Claret04}
        Claret, A. 2004, A\&A, 424, 919

\bibitem[2007]{Collet07}
        Collet, R., Asplund, M., \& Trampedach, R. 2007, A\&A, 649, 687

\bibitem[1970]{Connes70}
         Connes, P. 1970, \araa, 8, 209

\bibitem[1992]{Dearborn92} 
       Dearborn, D. S. P. 1992, Phys. Rep., 210, 367

\bibitem[2003]{Decin03}
        Decin, L., Vandenbussche, B., Waelkens, C., et al. 2003, 
        A\&A, 400, 709

\bibitem[1998]{Dyck98} 
       Dyck, H. M., van Belle, G. T., \& Thompson, R. R. 1998, 
       AJ, 116, 981  

\bibitem[2008]{Eggleton08} 
        Eggleton, P. P., Dearborn, D. S. P., \& Lattanzio, J. C. 2008,
        ApJ, 677, 581

\bibitem[1997]{ESA97}
        ESA. 1997, The Hipparcos and Tycho Catalogues, ESA SP-1200,
        Noordwijk 

\bibitem[1983]{Guelachivili83}
         Guelachivili, G., De Villeneuve, D., Farrenq, R., Urban, W.,
         \& Verges, J. 1983, J. Mol. Spectros., 98, 64

\bibitem[1975]{Gusafsson75}
        Gustafsson, B., Bell, R. A., Eriksson, K., \& Nordlund, \AA. 1975, 
        A\&A, 42, 407

\bibitem[1979]{Hall79}
        Hall, D. N. B., Ridgway, S. T., Bell, E.A., \& Yarborough,
	      J. M. 1979, Proc. Soc. Photo-Opt. Instrum. Eng., 172, 121

\bibitem[1984]{Harris84}
        Harris, M., \& Lambert, D. L. 1984,
        ApJ, 285, 674

\bibitem[1985]{Harris85} 
        Harris, M., Lambert, D. L., \& Smith, V. V. 1985, ApJ, 299, 375

\bibitem[1984]{Harris87}
        Harris, M., Lambert, D. L., Hinkle, K. H., Gustafsson, B., 
        \& Eriksson, K.  1987, ApJ, 316, 294

\bibitem[1984]{Hinkle76}
        Hinkle, K. H.,  Lambert, D. L., \& Snell, R. L.  1976, 
        ApJ, 210, 684

\bibitem[1999]{Jacquinet-Husson99} 
        Jacquinet-Husson, N., et al. 1999, \jqsrt, 62, 205

\bibitem[1979]{Kurucz79}
        Kurucz, R. L. 1979, ApJS, 40, 1

\bibitem[1984]{Lambert84}
        Lambert, D. L., Gustafsson, B., Eriksson, K., \& Hinkle, K. H.
        1984, ApJS, 62, 373

\bibitem[1993]{Langhoff93}
        Langhoff, S. R., \& Bauschlicher, C. W. 1993, 
        Chem. Phys. Lett., 211, 305

\bibitem[1991]{Lazaro91}
        Lazaro, C., Lynas-Gray, A. E., Clegg, R. E. S., et al.  1991, 
        \mnras, 249, 62

\bibitem[1981]{Lavas81}
        Lavas, F. J., Maki, A. G., \& Olson, W. B. 1981, 
        J. Mol. Spectros., 87, 449

\bibitem[1974]{Maillard74}
        Maillard, J. P. 1974, Highlights Astron., 3, 269

\bibitem[1999]{Matsuura99}  
        Matsuura, M., Yamamura, I., Murakami, H., Freund, M. M., \&
        Tanaka, M. 1999, A\&A, 348, 579

\bibitem[2002]{Mennesson02}
        Mennesson, B., Perrin, G., Chagnon, G., et al. 
        2002, ApJ, 579, 446 

\bibitem[1996]{Mihalas78}  
        Mihalas, D. 1978, Stellar Atmospheres, 2nd ed., 
        W. H. Freeman and Co., San Francisco

\bibitem[2005]{Mondal05}  
        Mondal, S., \& Chandrasekhar, T. 2005, AJ, 130, 842

\bibitem[1976]{Norton76}
        Norton, R. M.,\& Beer, R. 1976, J. Opt. Soc. America, 66, 259     

\bibitem[2004]{Ohnaka04}  
        Ohnaka, K. 2004, A\&A, 421, 1149

\bibitem[1996]{Ohnaka96}  
        Ohnaka, K., \& Tsuji, T. 1996, A\&A, 310, 933

\bibitem[1999]{Ohnaka99}  
        Ohnaka, K., \& Tsuji, T. 1999, A\&A, 345, 233

\bibitem[2000]{Ohnaka00}  
        Ohnaka, K., Tsuji, T., \& Aoki, W. 2000, A\&A, 353, 528

\bibitem[1997]{Partridge97}
        Partridge, H., \& Schwenke, D. W. 1997,  
        J. Chem. Phys., 106, 4618

 \bibitem[1998]{Perrin98}
        Perrin, G., Coud\'e du Foresto, Ridgway, S. T., et al. 
        1998, A\&A, 331, 619

\bibitem[2004]{Perrin04}
        Perrin, G., Ridgway, S. T., Coud\'e du Foresto,  et al. 
        2004, A\&A, 418, 675

\bibitem[1992]{Plez92a}
        Plez, B.,  1992, A\&AS, 94, 553

\bibitem[1992]{Plez92b}
        Plez, B., Brett, J. M., \& Nordlund, \AA. 1992, 
        A\&A, 256, 551

 \bibitem[2001]{Quirrenbach01}
         Quirrenbach, A. 2001, \araa, 39, 353

 \bibitem[1993]{Quirrenbach93}
         Quirrenbach, A., Mozrukewich, D., Armstrong, J. T., Buscher,
	 D. F., \& Hummel, C. A.  1993, ApJ, 406, 215

\bibitem[1997]{Reid97}
          Reid, M. J., \& Menten, K. M. 1997, ApJ, 476, 327

 \bibitem[1984]{Ridgway84}
        Ridgway, S. T., \&  Brault, J. W. 1984, \araa, 22, 291

 \bibitem[1980]{Ridgway80}
        Ridgway, S. T., Joyce, R. R., White, N. M., \&  Wing, R. F. 
        1980, ApJ, 235, 126

 \bibitem[1980]{Ryde02}
        Ryde, N., Lambert, D. L., Richter, M. J., \& Lacy, J. H. 2002,
        ApJ, 580, 447

\bibitem[1985]{Smith85}
        Smith, V. V., \& Lambert, D. L. 1985, ApJ, 294, 326

\bibitem[1986]{Smith86}
        Smith, V. V., \& Lambert, D. L. 1986, ApJ, 311, 843

\bibitem[1990]{Smith90}
        Smith, V. V., \& Lambert, D. L. 1990, ApJS, 72, 387

\bibitem[2007]{Suzuki07}
        Suzuki, T. K. 2007, ApJ, 659, 1592

\bibitem[1981]{Tipping81}
        Tipping, R. H., \& Chackerian, C., Jr. 1981, 
        J. Mol. Spectros., 88, 352

\bibitem[1978]{Tsuji78}  
        Tsuji, T. 1978, A\&A, 62, 29 

\bibitem[1981]{Tsuji81}  
        Tsuji, T. 1981, A\&A, 99, 48 

\bibitem[1986]{Tsuji86}  
        Tsuji, T. 1986, A\&A, 156, 8 (Paper I)

\bibitem[1988]{Tsuji88}  
        Tsuji, T. 1988, A\&A, 197, 185 (Paper II)

\bibitem[1991]{Tsuji91}
        Tsuji, T. 1991, A\&A, 245, 203 (Paper III)

\bibitem[2001]{Tsuji01}
        Tsuji, T. 2001, A\&A, 376, L1  

\bibitem[2002]{Tsuji02}
        Tsuji, T. 2002, ApJ, 575, 264   

\bibitem[2007]{Tsuji07}
        Tsuji, T. 2007, Proc. IAU Symp. 239 on ``Convecion in
        Asrophysics'', p. 307, Cambridge Univ. Press, Cambridge   

\bibitem[1997]{Tsuji97}
        Tsuji, T., Ohnaka, K., Aoki, W., \& Yamamura, I. 1997, A\&A,
	      320, L1  

\bibitem[1994]{Tsuji94}
        Tsuji, T., Ohnaka, K., Hinkle, K. H., \& Ridgway, S. T. 1994, 
        A\&A, 289, 469 

\bibitem[1955]{Unsold55}  
        Uns\"old, A. 1955, Physik der Sternatmosph\"aren mit
        Besonderer Ber\"ucksichigung der Sonne,  2ten Auf., 
        Springer, Berlin

\bibitem[1999]{vanBelle99}
        van Belle, G. T., Lane, B. F., Thompson, R. R., et al. 1999,
        AJ, 117, 521

\bibitem[2006]{Wahlin06}
         Wahlin, R., Eriksson, K., Gustafsson, B., et al. 2006, 
         Mem. S. A. It., 77, 955

\bibitem[1994]{Wallerstein94}
        Wallerstein, G., \& Morell, O. 1994, 
        A\&A, 281, L37 

\bibitem[2003]{Weiner03}
        Weiner, J., Hale, D. D. S., \& Townes, C. H. 2003, 
        ApJ, 589, 976 

\end{thebibliography}
\end{document}